\documentclass[prd,preprint,superscriptaddress,tightenlines,nofootinbib,
  eqsecnum,showpacs]{revtex4}

\usepackage{amsmath}
\usepackage{amsfonts}
\usepackage{amssymb}
\usepackage{bm}
\usepackage{hyperref}
\usepackage{mathrsfs}
\usepackage{graphicx}

\usepackage{ulem}
\usepackage[usenames]{color}

\allowdisplaybreaks

\DeclareSymbolFontAlphabet{\mathrsfs}{rsfs}
\DeclareMathAlphabet{\mathcal}{OMS}{cmsy}{m}{n}

\newcommand{\ud}{\mathrm{d}}
\newcommand{\ui}{\mathrm{i}}
\newcommand{\beq}{\begin{equation}}
\newcommand{\eeq}{\end{equation}}

\begin{document}

\title{Non-linear multipole interactions and gravitational-wave
  \\octupole modes for inspiralling compact binaries to
  third-and-a-half post-Newtonian order}

\author{Guillaume Faye}\email{faye@iap.fr}
\affiliation{$\mathcal{G}\mathbb{R}\varepsilon{\mathbb{C}}\mathcal{O}$,
  Institut d'Astrophysique de Paris --- UMR 7095 du CNRS,
  \\ Universit\'e Pierre \& Marie Curie, 98\textsuperscript{bis}
  boulevard Arago, 75014 Paris, France}

\author{Luc Blanchet}\email{blanchet@iap.fr}
\affiliation{$\mathcal{G}\mathbb{R}\varepsilon{\mathbb{C}}\mathcal{O}$,
  Institut d'Astrophysique de Paris --- UMR 7095 du CNRS,
  \\ Universit\'e Pierre \& Marie Curie, 98\textsuperscript{bis}
  boulevard Arago, 75014 Paris, France}

\author{Bala R. Iyer} \email{bri@rri.res.in} \affiliation{Raman
  Research Institute, Bangalore 560 080, India}

\date{\today}

\begin{abstract}
  This paper is motivated by the need to improve the post-Newtonian
  (PN) amplitude accuracy of waveforms for gravitational waves
  generated by inspiralling compact binaries, both for use in data
  analysis and in the comparison between post-Newtonian approximations
  and numerical relativity computations. It presents: (i) the
  non-linear couplings between multipole moments of general
  post-Newtonian matter sources up to order 3.5PN, including all
  contributions from tails, tails-of-tails and the non-linear memory
  effect; and (ii) the source mass-type octupole moment of
  (non-spinning) compact binaries up to order 3PN, which permits to
  complete the expressions of the octupole modes $(3,3)$ and $(3,1)$
  of the gravitational waveform to order 3.5PN. At this occasion we
  reconfirm by means of independent calculations our earlier results
  concerning the source mass-type quadrupole moment to order
  3PN. Related discussions on factorized resummed waveforms and the
  occurence of logarithmic contributions to high order are also
  included.
\end{abstract}


\maketitle

\section{Introduction} \label{sec:intro}

Coalescing compact binaries --- two neutron stars or black holes in
their late stage of evolution prior the final coalescence --- should
be the \textit{workhorse} source driving the network of advanced
interferometric gravitational-wave detectors on ground. The
post-Newtonian (PN) approximation is the appropriate technique to
extract accurate and reliable predictions from general relativity
theory for the inspiral phase of these systems. This constitutes the
starting point, in data analysis, to construct templates for double
neutron-star binaries and a crucial input to validate the early
inspiral phase of the numerical relativity waveforms for black-hole
binaries.

The non-linear evolution of the orbital phase due to gravitational
radiation reaction is the crucial ingredient in constructing these
templates. It has been completed for non-spinning compact binaries up
to order 3.5PN~\cite{BDIWW95, B98tail, BIJ02, BFIJ02, BI04mult,
  BDEI04, BDEI05dr}.\footnote{As usual the $n$PN order refers to the
  terms of order $1/c^{2n}$ in the waveform and energy flux, beyond
  the Einstein quadrupole formula which is referred to as the Newtonian
  approximation.} The amplitude of the signal, including all signal
harmonics besides the dominant one at twice the orbital frequency, has
been computed over the years with increasing precision and is now
complete to order 3PN~\cite{BIWW96, ABIQ04, KBI07, K07,
  BFIS08}. Furthermore the dominant quadrupole mode $(2,2)$ is also
known to order 3.5PN~\cite{FMBI12}. Our current program consists in
extending this computation and obtaining the full waveform up to order
3.5PN for all the modes $(\ell, m)$ in a spin-weighted
spherical-harmonic decomposition. In the present paper we shall, as
key milestones for this program:
\begin{enumerate}

\item Control all the non-linear couplings between multipole moments up to
  order 3.5PN for general matter sources; those couplings involve the
  important contributions of tails, tails-of-tails and the non-linear memory
  effect, as well as some extra contributions due to our specific definitions
  for the source multipole moments;

\item Obtain the source mass-type \textit{octupole} moment of (non-spinning)
  compact binaries up to order 3PN, which allows us to obtain the expressions
  of the octupole modes $(3,3)$ and $(3,1)$ of the waveform to order 3.5PN; we
  shall take this opportunity to recompute, using our new programs, the
  mass-type quadrupole moment to order 3PN and confirm the earlier
  results~\cite{BI04mult, BDEI04, BDEI05dr}.
\end{enumerate}
The full completion of our program will have to wait for the more difficult
computation of the source current-type quadrupole moment to order 3PN, which
is left for future work.

The plan of this paper is the following. Sec.~\ref{sec:MPM} is a
recapitulation of the basic definitions we use for source and
canonical multipole moments within the multipolar-post-Minkowskian
(MPM) formalism. In Sec.~\ref{sec:rad} we present (without proof) the
expressions of the radiative moments seen at infinity in terms of
canonical ones up to order 3.5PN for general matter sources (including
the various tail and memory effects), and the explicit links between
canonical and source moments. Next, Sec.~\ref{sec:octupole} deals with
the waveform of non-spinning compact binary sources. Notably, in
Subsec.~\ref{sec:octCM}, we compute the mass-type octupole moment for
general orbits to 3PN order, reduce it to the center-of-mass frame and
then to circular orbits, and, in Subsec.~\ref{sec:modes}, we obtain
the gravitational-wave modes $(3,3)$ and $(3,1)$ up to order 3.5PN for
circular orbits. Sec.~\ref{sec:factor} is devoted to the occurence of
logarithmic contributions in the MPM waveform to arbitrary high
non-linear orders, with an application to factorized resummed
waveforms in the effective one body (EOB) approach. The paper ends
with three more technical Appendices.

\section{Multipolar Post-Minkowskian expansion} 
\label{sec:MPM}

We look for the solution of the Einstein field equations in the vacuum
region outside the compact support of a general isolated matter
source. With
$h^{\alpha\beta}\equiv\sqrt{-g}g^{\alpha\beta}-\eta^{\alpha\beta}$
denoting the ``gothic'' metric deviation, where $g$ and
$g^{\alpha\beta}$ are respectively the determinant and the inverse of
the ``covariant'' metric $g_{\alpha\beta}$ and where
$\eta^{\alpha\beta}\equiv\text{diag}(-1,1,1,1)$ stands for the
Minkowski metric in Cartesian coordinates, the vacuum field equations
relaxed by the harmonic-gauge condition read
\begin{subequations}\label{EFE}
\begin{align}
\Box h^{\alpha\beta} &= \Lambda^{\alpha\beta}\bigl[h, \partial h,
  \partial^2h\bigr]\,,\label{EFEa}\\ \partial_\beta h^{\alpha\beta} &=
0\,.\label{EFEb}
\end{align}\end{subequations}
Here $\Box\equiv\eta^{\alpha\beta}\partial_\alpha\partial_\beta$
denotes the flat d'Alembertian operator, whereas the non-linear
gravitational source term $\Lambda^{\alpha\beta}$ is an expression of
second-order (at least) in the space-time components
$h^{\gamma\delta}$, which is quadratic in the first space-time
derivatives symbolized by $\partial h$ and linear in the second
space-time derivatives $\partial^2h$.

The multipolar-post-Minkowskian (MPM) expansion~\cite{BD86} is an
algorithmic procedure for generating iteratively the most general
solution of the field equations~\eqref{EFE} in the form of a
post-Minkowskian (or non-linearity) expansion whose coefficients are
themselves given by a multipole expansion physically valid outside the
compact support of the source. The multipole expansion is parametrized
by certain multipole moments characterizing the matter source but
left, in a first stage, as some unspecified functions of
time. However, among these moments, the mass monopole $M$ as well as
the mass and current dipoles, $M_i$ and $S_i$ respectively, are
constrained to be constant or to vary linearly with time; they
represent the ADM conserved mass, linear momentum and total angular
momentum of the source. In this paper we shall work in a mass-centred
frame such that $M_i=0$. Furthermore, an important assumption of the
MPM formalism is the stationarity in the past, namely the fact that
the matter source has been stationary in the remote past, before some
given date $-\mathcal{T}$. Thus, all multipole moments we shall
consider are assumed to be constant when $t\leqslant -\mathcal{T}$.

The starting point of the MPM algorithm is Thorne's~\cite{Th80}
linearized vacuum solution parametrized by two types of multipole
moments, called the \textit{source} moments: the mass-type moments
$I_L(t)$ and the current-type moments $J_L(t)$; they are such that
$I=M$, $I_i=M_i=0$ and $J_i=S_i$ are constant. Such general linearized
solution, referred to as ``canonical'', reads\footnote{Our notation is
  as follows. The retarded time is denoted as $t_r\equiv t-r/c$. The
  $n$-th time derivatives of multipole moments are indicated by
  superscripts $(n)$. $L = i_1 \cdots i_\ell$ denotes a multi-index
  composed of $\ell$ spatial indices (ranging from 1 to 3); $aL-1=a
  i_1 \cdots i_{\ell-1}$ and so on; $\partial_L = \partial_{i_1}
  \cdots \partial_{i_\ell}$ is the ``product'' of $\ell$ partial
  derivatives $\partial_i \equiv \partial / \partial x^i$; similarly
  $x_L = x_{i_1} \cdots x_{i_\ell}$ with $x_i$ being the spatial
  position, and $n_L = n_{i_1} \cdots n_{i_\ell}$ with
  $n_i=x_i/r$. Symmetrization over indices is denoted by
  $T_{(ij)}=\frac{1}{2}(T_{ij}+T_{ji})$. The symmetric-trace-free
  (STF) projection is indicated with a hat, \textit{i.e.} $\hat{n}_L
  \equiv \text{STF}[n_L]$, or by angular brackets $\langle\rangle$
  surrounding the relevant indices, \textit{e.g.}
  $\hat{n}_{ijk}=n_{\langle
    ijk\rangle}=n_in_jn_k-\frac{1}{5}[\delta_{ij}n_k +\delta_{jk}n_i
    +\delta_{ki}n_j]$. Underlined indices mean that they should be
  excluded from the STF projection, \textit{e.g.} $T_{\langle
    i\underline{a}j\rangle}=\frac{1}{2}(T_{iaj}+T_{jai}) -
  \frac{1}{3}\delta_{ij}T_{kak}$. The multipole moments we use,
  $\{I_L, J_L, W_L, X_L, Y_L, Z_L\}$, $\{M_L, S_L\}$ and $\{U_L,
  V_L\}$, are all STF, hence \textit{e.g.} $I_L=\hat{I}_L=I_{\langle
    L\rangle}$. In the case of summed-up multi-indices $L$, we do not
  write the $\ell$ summations from 1 to 3 over the dummy indices. The
  Levi-Civita antisymmetric symbol is denoted $\varepsilon_{iab}$
  (with $\varepsilon_{123}=1$).}
\begin{subequations} \label{hcan1}
\begin{align}
h^{00}_{\mathrm{can}\,(1)} &= -\frac{4}{c^2}\sum_{\ell = 0}^{+\infty}
\frac{(-)^\ell}{\ell !} \partial_L \left[ r^{-1} I_L (t_r)\right] \,
,\\ h^{0i}_{\mathrm{can}\,(1)} &= \frac{4}{c^3}\sum_{\ell =
  1}^{+\infty} \frac{(-)^\ell}{\ell!}  \left\{ \partial_{L-1} \left[
  r^{-1} I_{iL-1}^{(1)} (t_r)\right] + \frac{\ell}{\ell+1}
\varepsilon_{iab} \, \partial_{aL-1} \left[ r^{-1} J_{bL-1}
  (t_r)\right]\right\} \, ,\\ h^{ij}_{\mathrm{can}\,(1)} &=
-\frac{4}{c^4} \sum_{\ell = 2}^{+\infty} \frac{(-)^\ell}{\ell !}
\left\{ \partial_{L-2} \left[ r^{-1} I_{ijL-2}^{(2)} (t_r)\right] +
\frac{2\ell}{\ell+1} \partial_{aL-2} \left[ r^{-1} \varepsilon_{ab(i}
  J_{j)bL-2}^{(1)} (t_r)\right]\right\}\,.
\end{align}
\end{subequations}
It satisfies the relaxed linearized vacuum field equations $\Box
h_{\mathrm{can}\,(1)}^{\alpha\beta}=0$ and the harmonic gauge
condition $\partial_\beta h_{\mathrm{can}\,(1)}^{\alpha\beta} =0$,
formally at any point but $r=0$. However this solution is not the most
general one, as we can always perform an arbitrary linearized gauge
transformation maintaining the harmonic-gauge condition. Introducing
an arbitrary gauge vector $\varphi_{(1)}^\alpha$ satisfying
$\Box\varphi_{(1)}^\alpha=0$ (except at $r=0$), which will be
parametrized by four supplementary types of (unconstrained) multipole
moments $W_L(t)$, $X_L(t)$, $Y_L(t)$ and $Z_L(t)$ called the
\textit{gauge} moments, we can write
\begin{subequations} \label{phi1}
\begin{align}
\varphi^0_{(1)} =& \frac{4}{c^3} \sum_{\ell = 0}^{+\infty}
\frac{(-)^\ell}{\ell !}  \partial_L \left[ r^{-1} W_L (t_r)\right]
\,, \\ \varphi^i_{(1)} =& -\frac{4}{c^4} \sum_{\ell = 0}^{+\infty}
\frac{(-)^\ell}{ \ell !}  \partial_{iL} \left[ r^{-1} X_L
  (t_r)\right] \nonumber \\ & -\frac{4}{c^4} \sum_{\ell = 1}^{+\infty}
\frac{(-)^\ell}{\ell !}  \left\{ \partial_{L-1} \left[ r^{-1} Y_{iL-1}
  (t_r)\right] + \frac{\ell}{\ell+1} \varepsilon_{iab} \,
\partial_{aL-1} \left[ r^{-1} Z_{bL-1} (t_r)\right]\right\} \,.
\end{align}
\end{subequations}
The linear gauge terms take the form
$\partial\varphi_{(1)}^{\alpha\beta} \equiv
\partial^\alpha\varphi_{(1)}^{\beta} +
\partial^\beta\varphi_{(1)}^{\alpha} -
\eta^{\alpha\beta}\partial_\gamma\varphi_{(1)}^{\gamma}$ so that the
most general linearized vacuum solution in harmonic coordinates reads
\begin{equation} \label{hgen1}
h_{\mathrm{gen}\,(1)}^{\alpha\beta} =
h_{\mathrm{can}\,(1)}^{\alpha\beta}\bigl[I_L,J_L\bigr] +
\partial\varphi_{(1)}^{\alpha\beta}\bigl[W_L,X_L,Y_L,Z_L\bigr]\,.
\end{equation}
Starting from $h_{\mathrm{gen}\,(1)}$ the MPM algorithm will generate a full
post-Minkowskian solution of the field equations~\eqref{EFE}, \textit{i.e.} a
solution given as a formal non-linear expansion series in powers of Newton's
constant $G$, as shown in Eq.~\eqref{PMgen} below. Suppose that one has
succeeded in generating all the post-Minkowskian coefficients up to some order
$n-1$, say $h_{\mathrm{gen}\,(2)}$, $\cdots$, $h_{\mathrm{gen}\,(n-1)}$. Then
the precise procedure by which the next post-Minkowskian coefficient,
\textit{i.e.} $h_{\mathrm{gen}\,(n)}$, is generated is as follows~\cite{BD86}.
One decomposes this coefficient into two terms,
\begin{equation} \label{hgenn}
h^{\alpha\beta}_{\mathrm{gen}\,(n)} =
u^{\alpha\beta}_{\mathrm{gen}\,(n)} +
v^{\alpha\beta}_{\mathrm{gen}\,(n)} \,.
\end{equation}
The first one is defined as the standard (flat) retarded integral,
denoted $\Box^{-1}_\mathrm{ret}$, of the iterated source term coming
from the relaxed Einstein field equation~\eqref{EFEa}. Namely, after
obtaining from the previous iterations the $n$-th post-Minkowskian
order source term as some
$\Lambda_{(n)}=\Lambda_{(n)}[h_{\mathrm{gen}\,(1)}, \cdots,
  h_{\mathrm{gen}\,(n-1)}]$, we pose
\begin{equation} \label{un}
u^{\alpha\beta}_{\mathrm{gen}\,(n)} = \mathop{\mathrm{FP}}_{B=0} \,
\Box^{-1}_\mathrm{ret} \left[ \widetilde{r}^B
  \Lambda_{(n)}^{\alpha\beta} \right] \,.
\end{equation}
Crucial to the MPM algorithm is the regularization process based on
analytic continuation in a complex parameter $B$ which enters a
regulator factor,
\begin{equation} \label{regulator}
\widetilde{r}^B \equiv \left(\frac{r}{r_0}\right)^B \, ,
\end{equation}
multiplying the source term. Here $r_0$ is an arbitrary constant
length scale. The regulator~\eqref{regulator} permits, thanks to
analytic continuation, to cure the divergency of the multipole
expansion when $r\to 0$ that follows from the fact that the vacuum
solution is physically valid only outside the matter source and is yet
to be matched to the actual solution inside it.\footnote{The matching
  to a general isolated post-Newtonian matter source in the external
  near zone of this source has been elucidated within this
    formalism in Refs.~\cite{B95, B98mult, PB02, BFN05}.} Finally, an
operation of taking the finite part (FP), \textit{i.e.} picking up the
term with zeroth power of $B$ in the Laurent expansion of the
expression when $B\to 0$, is applied. This fully defines the
expression~\eqref{un} as a particular solution of $\Box
u_{\mathrm{gen}\,(n)} = \Lambda_{\mathrm{gen}\,(n)}$ everywhere except
at $r=0$.

The second term in Eq.~\eqref{hgenn} ensures that the harmonic gauge
condition $\partial_\beta h^{\alpha\beta}_{\mathrm{gen}\,(n)}=0$ is
satisfied. It is algorithmically computed from the divergence of the
first term, namely $w^{\alpha}_{\mathrm{gen}\,(n)} \equiv
\partial_{\beta}u^{\alpha\beta}_{\mathrm{gen}\,(n)}$, which is necessarily a
retarded solution of the source-free d'Alembertian equation, $\Box
w^{\alpha}_{\mathrm{gen}\,(n)}=0$. That solution can thus always be written as
\begin{subequations} \label{wn}
\begin{align}
w^0_{\mathrm{gen}\,(n)} &= \sum_{\ell = 0}^{+\infty} \partial_L
\left[r^{-1} N_L(t_r)\right] \,, \\ w^i_{\mathrm{gen}\,(n)} & =
\sum_{\ell =0}^{+\infty}\partial_{iL} \left[ r^{-1} P_L (t_r)
  \right] \nonumber \\ & + \sum_{\ell = 1}^{+\infty} \Bigl\{
\partial_{L-1} \left[ r^{-1} Q_{iL-1} (t_r) \right] +
\varepsilon_{iab} \, \partial_{aL-1} \left[r^{-1} R_{bL-1} (t_r)
  \right] \Bigr\} \,,
\end{align}
\end{subequations}
where the STF multipole moments $\{N_L, P_L, Q_L, R_L\}$ are given by
some (very complicated at high post-Minkowskian orders $n$)
functionals of the initial source and gauge moments $\{I_L, J_L, W_L,
X_L, Y_L, Z_L\}$. We then pose~\cite{BD86,B98mult} 
\begin{subequations} \label{vn}
\begin{align}
v^{00}_{\mathrm{gen}\,(n)} &= - c\, r^{-1} N^{(-1)} + \partial_a \left[
  r^{-1} \left(- c\, N^{(-1)}_a+ c^2 Q^{(-2)}_a -3P_a\right) \right] \, ,
\\ v^{0i}_{\mathrm{gen}\,(n)} &= r^{-1} \left( - c\, Q^{(-1)}_i +3 c^{-1}
P^{(1)}_i\right) - \varepsilon_{iab} \, \partial_a \left[ r^{-1} c\, 
  R^{(-1)}_b \right] - \sum_{\ell = 2}^{+\infty}\partial_{L-1} \left[
  r^{-1} N_{iL-1} \right] \, , \\ v^{ij}_{\mathrm{gen}\,(n)} &= -
\delta_{ij} r^{-1} P + \sum_{\ell = 2}^{+\infty} \biggl\{ 2
\delta_{ij}\partial_{L-1} \left[ r^{-1} P_{L-1}\right] - 6
\partial_{L-2(i} \left[ r^{-1} P_{j)L-2}\right] \nonumber \\ & \quad +
\partial_{L-2} \left[ r^{-1} (c^{-1}N^{(1)}_{ijL-2} + 3 c^{-2} P^{(2)}_{ijL-2} -
  Q_{ijL-2}) \right] - 2 \partial_{aL-2}\left[ r^{-1}
  \varepsilon_{ab(i} R_{j)bL-2} \right] \biggr\} \,.
\end{align}
\end{subequations}
It can readily be checked that $\partial_\beta
v^{\alpha\beta}_{\mathrm{gen}\,(n)}=-w^{\alpha}_{\mathrm{gen}\,(n)}$,
hence $\partial_\beta h^{\alpha\beta}_{\mathrm{gen}\,(n)}=0$. Since we
also have $\Box v^{\alpha\beta}_{\mathrm{gen}\,(n)}=0$, we see that
the $n$-th post-Minkowskian order piece of the gravitational
field~\eqref{hgenn} satisfies the relaxed field equations in harmonic
coordinates at order $n$.  Note the presence in Eqs.~\eqref{vn} of
anti-derivatives, denoted \textit{e.g.} $N^{(-1)}$, which are
associated with the secular losses of energy, linear momentum and
angular momentum of the source through gravitational radiation.

Finally, we get a full solution of the vacuum Einstein field
equations~\eqref{EFE}, parametrized by two sets of source moments
$I_L$, $J_L$ and four sets of gauge moments $W_L$, $X_L$, $Y_L$,
$Z_L$, in the form of the post-Minkowskian expansion series
\begin{equation} \label{PMgen}
h_\text{gen}^{\alpha\beta} = \sum_{n=1}^{+\infty} G^n
h_{\mathrm{gen}\,(n)}^{\alpha\beta}\bigl[I_L,J_L,W_L,X_L,Y_L,Z_L\bigr]\,.
\end{equation}
It was proved~\cite{BD86} that this represents physically the most
general solution of the vacuum field equations outside an isolated
matter system.  Thanks to the matching, all the multipole moments
therein have been given explicit closed-form expressions as integrals
over the matter and gravitational fields of a general post-Newtonian
source~\cite{B98mult, PB02}.

The explicit MPM construction leading to Eq.~\eqref{PMgen} is quite
complicated in practice but now entirely performed on a computer.\footnote{The
  MPM algorithm is implemented by using the algebraic computing software
  \textit{Mathematica} together with the tensor package
  \textit{xAct}~\cite{xtensor}. Intermediate results along with the source
  codes can be provided by the authors on request.} It is often convenient to
simplify it by considering, instead of the six sets of source and gauge
moments, only two, called the \textit{canonical} mass-type and current-type
multipole moments, $M_L(t)$ and $S_L(t)$ respectively. Indeed it has been
proved~\cite{Th80, BD86} that the most general solution is actually
parametrized by two and only two sets of moments --- by definition these
canonical $M_L$ and $S_L$ moments. The simplest MPM construction, here
referred to as ``canonical'', is obtained by annulling all the gauge moments
in Eq.~\eqref{PMgen} and starting with $M_L$, $S_L$ in place of $I_L$, $J_L$,
\textit{i.e.}
\begin{equation} \label{PMcan}
h_\text{can}^{\alpha\beta} = \sum_{n=1}^{+\infty} G^n
h_{\mathrm{gen}\,(n)}^{\alpha\beta}\bigl[M_L,S_L,0,0,0,0\bigr]\,.
\end{equation}
This means that the iteration now begins at linearized order with the
solution $h_{\mathrm{can}\,(1)}[M_L,S_L]$. However, even if we proceed
with the simpler construction~\eqref{PMcan}, we still have to relate
the canonical moments $\{M_L, S_L\}$ to the source and gauge moments
$\{I_L, J_L, W_L, X_L, Y_L, Z_L\}$, because only the latter are known
as explicit integrals over the matter and gravitational fields of the
source. To relate these two sets, we impose that the two
constructions~\eqref{PMgen} and~\eqref{PMcan} are to be
\textit{isometric}, \textit{i.e.} to differ by a coordinate
transformation. It can be shown --- see notably Ref.~\cite{BFIS08} for
an explicit derivation at quadratic order --- that this yields unique
relations of the type
\begin{subequations}\label{cangen}
\begin{align}
M_L &= I_L + \mathcal{M}_L\left[I, J, W, X, Y, Z\right]\,,\\ S_L &=
J_L + \mathcal{S}_L\left[I, J, W, X, Y, Z\right]\,,
\end{align}
\end{subequations}
where $\mathcal{M}_L$ and $\mathcal{S}_L$ denote some non-linear
functionals of the source and gauge moments that are at least
quadratic and start only at the high order 2.5PN. When the
relations~\eqref{cangen} are satisfied, the two sets of moments
$\{M_L, S_L\}$ and $\{I_L, J_L, W_L, X_L, Y_L, Z_L\}$ describe the
same physical matter source. We shall give in Sec.~\ref{sec:cansource}
below their most up-to-date explicit forms.

\section{The radiative multipole moments}
\label{sec:rad}

In the previous section we reviewed the MPM
solutions~\eqref{PMgen}--\eqref{PMcan}, which are valid all-over the
exterior of the source, in particular at future null
infinity. However, these solutions exhibit a logarithmic far-zone
structure $\sim (\ln r)^p/r^k$ when expanded as $r\to +\infty$ with
$t_r\equiv t-r/c=\mathrm{const}$, where $t$ and $r$ refer to the
harmonic coordinates (see also Sec.~\ref{sec:resum}). This is due to
the well-known logarithmic deviation of the null cones with respect to
the retarded cones $t-r/c$ in this coordinate grid. It is thus
convenient to introduce so-called \textit{radiative} coordinates $(T,
R)$ such that $T_R\equiv T-R/c$ is a null coordinate, or becomes
asymptotically null in the limit $R\to +\infty$. We then have (with
the angular coordinates being untouched)
\begin{equation}\label{TRtr}
T_R = t_r -\frac{2 G
  M}{c^3}\ln\left(\frac{r}{c b}\right) +
\mathcal{O}\left(\frac{1}{r}\right)\,,
\end{equation}
where $M$ is the total mass of the source and $b$ is an arbitrary
constant time scale, \textit{a priori} unrelated to the constant $r_0$
introduced in the MPM regulator~\eqref{regulator}. In radiative
coordinates the structure of the expansion when $R\to +\infty$ with
$T_R=\mathrm{const}$ is merely $\sim 1/R^k$~\cite{B87}.

The STF radiative moments $\{U_L, V_L\}$ are then defined from the
leading $1/R$ term of the asymptotic waveform by~\cite{Th80}
\begin{align} \label{gijTT}
g_{ij}^\text{TT} = \delta_{ij} &+ \frac{4G}{c^2R}
\left[\sum^{+\infty}_{\ell=2}\frac{1}{c^\ell \ell !} \biggl\{ N_{L-2}
  \, U_{ijL-2}(T_R) - \frac{2\ell}{c(\ell+1)} \, N_{aL-2}
  \,\varepsilon_{ab(i} \, V_{j)bL-2}(T_R)\biggr\}\right]^\text{TT}
\nonumber\\ & + \mathcal{O}\left(\frac{1}{R^2}\right)\,,
\end{align}
where the superscript TT refers to the usual algebraic
transverse-traceless projection. Below we shall present (without the
full derivations) the expressions of the radiative multipole moments
needed to control the waveform up to order 3.5PN. These results are
obtained by implementing the MPM algorithm reviewed in the previous
section.

Our goal being to obtain the radiative moments $\{U_L, V_L\}$ as
functionals of the source and gauge moments $\{I_L, J_L, W_L, X_L,
Y_L, Z_L\}$, it is useful to know beforehand which types of
interactions between any moments $A_J$ and $B_K$ (with $j$ and $k$
indices respectively), among the set of source and gauge moments, are
allowed in a given radiative mass moment $U_L$ or current moment $V_L$
up the 3.5PN order. To answer that question in the case of quadratic
interactions, say $A_J\times B_K$ where $A_J, B_K \in \{I_L, J_L, W_L,
X_L, Y_L, Z_L\}$, we have developed some ``selection rules'' following
Refs.~\cite{BFIS08, FMBI12}. The interactions allowed by those rules
at order 3.5PN are given in Table~\ref{tab:rules}.

The two panels of Table~\ref{tab:rules} show the maximal number of
indices $\ell_\text{max}[A_J, B_K]$ on the mass moment $U_L$ and the
current moment $V_{L}$ (respectively), beyond which $U_L$ or $V_L$
cannot contain products of the two multipole moments $A_J$ and $B_K$
or their time derivatives (or possibly time anti-derivatives) at the
3.5PN order in the waveform. Thus, when $\ell>\ell_\text{max}[A_J,
  B_K]$ we can safely ignore the multipole interaction $A_J\times B_K$
in $U_L$ or $V_L$ since it will be of higher PN order. When
$\ell\leqslant\ell_\text{max}[A_J, B_K]$ we can deduce all the
possible relevant interactions $A_J\times B_K$ by noticing that
$j+k=\ell_\text{max}[A_J, B_K]$ if the product $A_J B_K$ has the same
parity as the radiative moment containing the interaction
(\textit{i.e.} both $A_J$ and $B_K$ are mass moments or both are
current moments in $U_L$; one is a mass moment and the other is a
current moment in $V_L$), and $j+k=\ell_\text{max}[A_J, B_K]+1$ in all
other cases.\footnote{From Eqs.~\eqref{phi1} we see that the gauge
  moments $W_L$, $X_L$ and $Y_L$ have the same parity as the mass
  moment $I_L$, while the gauge moment $Z_L$ has the same parity as
  the current moment $J_L$.} Finally, the case $\ell_\text{max}[A_J,
  B_K]<2$ is obviously impossible since radiative moments have at
least $\ell=2$. This case is indicated by dashes in the two panels of
Table~\ref{tab:rules}. We emphasize that the latter rules apply to
quadratic interactions, which are the most tricky to control
thoroughly. In the present paper we shall also need to include some
cubic interactions. These are simpler to look for and will be dealt
with separately.
\begin{table}[t]
\begin{center}
\begin{tabular}{|l||c|c|c|c|c|c|} 
\hline
$A_J$ & \multicolumn{6}{c|}{$B_K$} \\ \cline{2-7} 
      &$I_K$&$J_K$&$W_K$&$X_K$&$Y_K$&$Z_K$ \\ \hline 
$I_J$ &  6  &  4  &  4  &  2 &  4  &  2  \\ \hline
$J_J$ &  4  &  4  &  2  & -- &  2  &  2  \\ \hline
$W_J$ &  4  &  2  &  2  & -- &  2  & --  \\ \hline
$X_J$ &  2  & --  &  -- & -- & --  & --  \\ \hline
$Y_J$ &  4  &  2  &  2  & -- &  2  & --  \\ \hline
$Z_J$ &  2  &  2  &  -- & -- & --  & --  \\ \hline
\end{tabular}
\hspace{2cm}
\begin{tabular}{|l||c|c|c|c|c|c|} 
\hline
$A_J$ & \multicolumn{6}{c|}{$B_K$} \\ \cline{2-7} 
      &$I_K$&$J_K$&$W_K$&$X_K$&$Y_K$&$Z_K$ \\ \hline 
$I_J$ &  5  &  5  &  3  &  -- &  3  &  3  \\ \hline
$J_J$ &  5  &  3  &  3  & -- &  3  &  --  \\ \hline
$W_J$ &  3  &  3  &  --  & -- &  --  & --  \\ \hline
$X_J$ &  --  &  --  &  -- & -- & --  & --  \\ \hline
$Y_J$ &  3  &  3  &  --  & -- &  --  & --  \\ \hline
$Z_J$ &  3  &  --  &  -- & -- & --  & --  \\ \hline
\end{tabular}		
\caption{Left panel: Values of $\ell_\text{max}[A_J,B_K]$ for the mass
  multipole moment $U_L$ at the 3.5PN order for the various possible
  choices of multipole interactions between $A_J$ and $B_K \in \{I, J,
  W, X, Y, Z\}$. Right panel: likewise but for the current multipole
  moment $V_L$. We must have $j+k=\ell_\text{max}[A_J, B_K]$ if the
  product $A_J B_K$ has the same parity as the radiative moment
  containing the interaction $A_J\times B_K$, \textit{i.e.} both $A_J$
  and $B_K$ belong to $\{I, W, X, Y\}$ or both belong to $\{J, Z\}$ in
  $U_L$, $A_J$ belongs to $\{I, W, X, Y\}$ and $B_K$ belongs to $\{J,
  Z\}$ or \textit{vice-versa} in $V_L$; in all other cases
  $j+k=\ell_\text{max}[A_J, B_K]+1$. Impossible (because too low)
  values of $\ell_\text{max}[A_J,B_K]$ are indicated by
  dashes.}\label{tab:rules}
\end{center}
\end{table}

\subsection{Radiative moments in terms of canonical moments}
\label{sec:radcan}

Like in our previous papers~\cite{BFIS08, FMBI12}, in order to
simplify the presentation, we shall first present the radiative
moments $\{U_L, V_L\}$ in terms of the canonical moments $\{M_L,
S_L\}$, and only in a second stage shall we give the canonical moments
in terms of the set of source and gauge moments $\{I_L, J_L, W_L, X_L,
Y_L, Z_L\}$. Clearly, the selection rules provided in
Table~\ref{tab:rules} apply to the full set of quadratic interactions
with $A_J, B_K \in \{I_L, J_L, W_L, X_L, Y_L, Z_L\}$ as well as to the
restricted set with $A_J, B_K \in \{M_L, S_L\}$.

To display the results, it is also convenient to group the terms in
the radiative moments $U_L$ and $V_L$ into different types, namely on
the one hand all the instantaneous terms, and on the other hand the
hereditary terms~\cite{BD92} which comprise the tails, the
tails-of-tails, and the non-linear memory integrals:
\begin{subequations}\label{UVL}
\begin{align}
U_L &= U_L^\text{inst} + U_L^\text{tail} + U_L^\text{tail-tail} +
U_L^\text{mem} + \delta U_L\,,\\ V_L &= V_L^\text{inst} +
V_L^\text{tail} + V_L^\text{tail-tail} + V_L^\text{mem} + \delta
  V_L\,.
\end{align}\end{subequations}
The terms $\delta U_L$ and $\delta V_L$ in the above decomposition represent
the contributions occuring at 4PN or higher orders in the waveform (generally
with a more complex non-linear structure), which will be neglected here. Below
we shall show the formulas needed to complete the waveform including all its
relevant harmonics up to order 3.5PN. Some of those formulas have already been
partially published in Refs.~\cite{BFIS08, FMBI12}, but we reproduce them all
in order to be  self-contained for the convenience of the user.

\subsubsection{Instantaneous terms}
\label{sec:inst}

These terms, in which by definition all the canonical moments are
evaluated at the current (retarded) time $T_R=T-R/c$, are the most
intricate terms to obtain. For the mass-type moments they are given
by:
\begin{subequations}
\begin{align}
U_{ij}^\text{inst} &= M^{(2)}_{ij} \nonumber \\ &+\frac{G}{
  c^5}\biggl[ \frac{1}{ 7}M^{(5)}_{a\langle i}M_{j\rangle a} -
  \frac{5}{7} M^{(4)}_{a\langle i}M^{(1)}_{j\rangle a} -\frac{2}{7}
  M^{(3)}_{a\langle i}M^{(2)}_{j\rangle a}
  +\frac{1}{3}\varepsilon_{ab\langle i}M^{(4)}_{j\rangle
    a}S_{b}\biggr]\nonumber \\ & + \frac{G}{c^7} \bigg[- \frac{1}{432}
  M_{ab} M_{ijab}^{(7)} + \frac{1}{432} M_{ab}^{(1)} M_{ijab}^{(6)} -
  \frac{5}{756} M_{ab}^{(2)} M_{ijab}^{(5)} + \frac{19}{648}
  M_{ab}^{(3)} M_{ijab}^{(4)} \nonumber \\ & \quad\qquad +
  \frac{1957}{3024} M_{ab}^{(4)} M_{ijab}^{(3)} + \frac{1685}{1008}
  M_{ab}^{(5)} M_{ijab}^{(2)} + \frac{41}{28} M_{ab}^{(6)}
  M_{ijab}^{(1)} + \frac{91}{216} M_{ab}^{(7)} M_{ijab} \nonumber \\ &
  \quad\qquad - \frac{5}{252} M_{ab \langle i} M_{j \rangle ab}^{(7)}
  + \frac{5}{189} M_{ab \langle i}^{(1)} M_{j \rangle ab}^{(6)} +
  \frac{5}{126} M_{ab \langle i}^{(2)} M_{j \rangle ab}^{(5)} +
  \frac{5}{2268} M_{ab \langle i}^{(3)} M_{j \rangle ab}^{(4)}
  \nonumber \\ & \quad\qquad + \frac{5}{42} S_a S_{ija}^{(5)} +
  \frac{80}{63} S_{a \langle i} S_{j \rangle a}^{(5)} + \frac{16}{63}
  S_{a \langle i}^{(1)} S_{j \rangle a}^{(4)} - \frac{64}{63} S_{a
    \langle i}^{(2)} S_{j \rangle a}^{(3)} \nonumber \\ & \quad\qquad
  + \varepsilon_{ac \langle i} \Big( \frac{1}{168} S_{j \rangle
    bc}^{(6)} M_{ab} + \frac{1}{24} S_{j\rangle bc}^{(5)} M_{ab}^{(1)}
  + \frac{1}{28} S_{j \rangle bc}^{(4)} M_{ab}^{(2)} + \frac{3}{56}
  S_{j \rangle bc}^{(2)} M_{ab}^{(4)} \nonumber \\ & \quad\qquad\qquad
  + \frac{187}{168} S_{j \rangle bc}^{(1)} M_{ab}^{(5)} +
  \frac{65}{84} S_{j \rangle bc} M_{ab}^{(6)} + \frac{1}{189} M_{j
    \rangle bc}^{(6)} S_{ab} - \frac{1}{189} M_{j \rangle bc}^{(5)}
  S_{ab}^{(1)} \nonumber \\ & \quad\qquad\qquad + \frac{10}{189} M_{j
    \rangle bc}^{(4)} S_{ab}^{(2)} + \frac{32}{189} M_{j \rangle
    bc}^{(3)} S_{ab}^{(3)} + \frac{65}{189} M_{j \rangle bc}^{(2)}
  S_{ab}^{(4)} - \frac{5}{189} M_{j \rangle bc}^{(1)} S_{ab}^{(5)}
  \nonumber\\ & \quad\qquad\qquad - \frac{10}{63} M_{j \rangle bc}
  S_{ab}^{(6)} - \frac{1}{6} S_{j \rangle bc}^{(3)} M_{ab}^{(3)} \Big)
  \bigg] \,,\\
U_{ijk}^\text{inst} &= M^{(3)}_{ijk} \nonumber \\ & +{G\over
  c^5}\bigg[-{4\over3}M^{(3)}_{a\langle i}M^{(3)}_{jk\rangle
    a}-{9\over4}M^{(4)}_{a\langle i}M^{(2)}_{jk\rangle a} +
  {1\over4}M^{(2)}_{a\langle i}M^{(4)}_{jk\rangle a} -
  {3\over4}M^{(5)}_{a\langle i}M^{(1)}_{jk\rangle a} \nonumber\\ &
  \quad\qquad +{1\over4}M^{(1)}_{a\langle i}M^{(5)}_{jk\rangle a} +
              {1\over12}M^{(6)}_{a\langle i}M_{jk\rangle a}
              +{1\over4}M_{a\langle i}M^{(6)}_{jk\rangle a}
              \nonumber\\ & \quad\qquad +
                          {1\over5}\varepsilon_{ab\langle i}\bigg(
                          -12S^{(2)}_{j\underline{a}}M^{(3)}_{k\rangle
                            b}-8M^{(2)}_{j\underline{a}}S^{(3)}_{k\rangle b}
                          -3S^{(1)}_{j\underline{a}}M^{(4)}_{k\rangle
                            b}\nonumber\\ & \quad\qquad\qquad
                          -27M^{(1)}_{j\underline{a}}S^{(4)}_{k\rangle
                            b}-S_{j\underline{a}}M^{(5)}_{k\rangle
                            b}-9M_{j\underline{a}}S^{(5)}_{k\rangle b}
                          -{9\over4}S_{\underline{a}}M^{(5)}_{jk\rangle b}\bigg)
                          +{12\over5}S_{\langle
                            i}S^{(4)}_{jk\rangle}\bigg]\,,\\
U_{ijkl}^\text{inst} &= M^{(4)}_{ijkl} \nonumber \\ &+ {G\over c^3}
\bigg[ -{21\over5}M^{(5)}_{\langle ij}M_{kl\rangle }- {63
    \over5}M^{(4)}_{\langle ij}M^{(1)}_{kl\rangle }-
  {102\over5}M^{(3)}_{\langle ij}M^{(2)}_{kl\rangle }\bigg] \nonumber
\\ &+\frac{G}{c^5} \bigg[ \frac{7}{55} M_{a \langle i} M_{jkl \rangle
    a}^{(7)} + \frac{7}{55} M_{a \langle i}^{(1)} M_{jkl \rangle
    a}^{(6)} + \frac{1}{25} M_{a \langle i}^{(2)} M_{jkl \rangle
    a}^{(5)} - \frac{28}{11} M_{a \langle i}^{(3)} M_{jkl \rangle
    a}^{(4)} \nonumber \\ & \quad\qquad - \frac{273}{55} M_{a \langle
    i}^{(4)} M_{jkl \rangle a}^{(3)} - \frac{203}{55} M_{a \langle
    i}^{(5)} M_{jkl \rangle a}^{(2)} - \frac{49}{55} M_{a \langle
    i}^{(6)} M_{jkl \rangle a}^{(1)} + \frac{14}{275} M_{a \langle
    i}^{(7)} M_{jkl \rangle a} \nonumber \\ & \quad\qquad +
  \frac{14}{33} M_{a \langle ij} M_{kl \rangle a}^{(7)} +
  \frac{37}{33} M_{a \langle ij}^{(1)} M_{kl \rangle a}^{(6)} +
  \frac{9}{11} M_{a \langle ij}^{(2)} M_{kl \rangle a}^{(5)} +
  \frac{8}{33} M_{a \langle ij}^{(3)} M_{kl \rangle a}^{(4)} +
  \frac{9}{5} S_{\langle i} S_{jkl \rangle}^{(5)} \nonumber \\ &
  \quad\qquad + \frac{16}{5} S_{\langle ij} S_{kl \rangle}^{(5)} +
  \frac{48}{5} S_{\langle ij}^{(1)} S_{kl \rangle}^{(4)} +
  \frac{32}{5} S_{\langle ij}^{(2)} S_{kl \rangle}^{(3)} \nonumber
  \\ & \quad\qquad + \varepsilon_{ab \langle i} \Big(- \frac{3}{5}
  M_{j\underline{a}} S_{kl \rangle b}^{(6)} - \frac{63}{25}
  M_{j\underline{a}}^{(1)} S_{kl \rangle b}^{(5)} + \frac{3}{5}
  M_{j\underline{a}}^{(2)} S_{kl \rangle b}^{(4)} + \frac{18}{5}
  M_{j\underline{a}}^{(3)} S_{kl \rangle b}^{(3)} \nonumber \\ &
  \quad\qquad\qquad + \frac{9}{5} M_{j\underline{a}}^{(4)} S_{kl
    \rangle b}^{(2)} + \frac{3}{5} M_{j\underline{a}}^{(5)} S_{kl
    \rangle b}^{(1)} + \frac{3}{25} M_{j\underline{a}}^{(6)} S_{kl
    \rangle b} - \frac{8}{15} S_{j\underline{a}} M_{kl \rangle
    b}^{(6)} - \frac{24}{25} S_{j\underline{a}}^{(1)} M_{kl \rangle
    b}^{(5)} \nonumber \\ & \quad\qquad\qquad - \frac{8}{5}
  S_{j\underline{a}}^{(2)} M_{kl \rangle b}^{(4)} + \frac{16}{3}
  S_{j\underline{a}}^{(3)} M_{kl \rangle b}^{(3)} + \frac{72}{5}
  S_{j\underline{a}}^{(4)} M_{kl \rangle b}^{(2)} + \frac{56}{5}
  S_{j\underline{a}}^{(5)} M_{kl \rangle b}^{(1)} \nonumber \\ &
  \quad\qquad\qquad + \frac{232}{75} S_{j\underline{a}}^{(6)} M_{kl
    \rangle b} + \frac{29}{75} M_{jkl \rangle a}^{(6)} S_b \Big)
  \bigg] \,,\\
U_{ijklm}^\text{inst} &= M^{(5)}_{ijklm} \nonumber \\ &+ {G\over
  c^3}\bigg[ -{710\over21}M^{(3)}_{\langle
    ij}M^{(3)}_{klm\rangle}-{265\over7}M^{(2)}_{\langle
    ijk}M^{(4)}_{lm\rangle} -{120\over7}M^{(2)}_{\langle
    ij}M^{(4)}_{klm\rangle}\nonumber\\ & \quad\qquad
  -{155\over7}M^{(1)}_{\langle
    ijk}M^{(5)}_{lm\rangle}-{41\over7}M^{(1)}_{\langle
    ij}M^{(5)}_{klm\rangle} -{34\over7}M_{\langle
    ijk}M^{(6)}_{lm\rangle}-{15\over7}M_{\langle
    ij}M^{(6)}_{klm\rangle}\bigg]\,,\label{U5}\\
U_{ijklmn}^\text{inst} &= M^{(6)}_{ijklmn} \nonumber \\ &+
\frac{G}{c^3} \bigg[ - \frac{45}{28} M_{\langle ij} M_{klmn
    \rangle}^{(7)} - \frac{111}{28} M_{\langle ij}^{(1)} M_{klmn
    \rangle}^{(6)} - \frac{561}{28} M_{\langle ij}^{(2)} M_{klmn
    \rangle}^{(5)} \nonumber \\ & \quad\qquad - \frac{1595}{28}
  M_{\langle ij}^{(3)} M_{klmn \rangle}^{(4)} - \frac{2505}{28}
  M_{\langle ij}^{(4)} M_{klmn \rangle}^{(3)} - \frac{2115}{28}
  M_{\langle ij}^{(5)} M_{klmn \rangle}^{(2)} \nonumber \\ &
  \quad\qquad - \frac{909}{28} M_{\langle ij}^{(6)} M_{klmn
    \rangle}^{(1)} - \frac{159}{28} M_{\langle ij}^{(7)} M_{klmn
    \rangle} - \frac{15}{7} M_{\langle ijk} M_{lmn \rangle}^{(7)} -
  \frac{75}{7} M_{\langle ijk}^{(1)} M_{lmn \rangle}^{(6)} \nonumber
  \\ & \quad\qquad - \frac{135}{7} M_{\langle ijk}^{(2)} M_{lmn
    \rangle}^{(5)} - \frac{505}{21} M_{\langle ijk}^{(3)} M_{lmn
    \rangle}^{(4)} \bigg] \,.
\end{align}
\end{subequations}
In the above expressions, the $1/c^5$ terms in $U_{ijkl}^\text{inst}$
and $1/c^3$ terms in $U_{ijklmn}^\text{inst}$ are new with the present
paper; the other terms were obtained in
Refs.~\cite{BFIS08,FMBI12}. For the current-type moments we have:
\begin{subequations}
\begin{align}
  V_{ij}^\text{inst} &= S^{(2)}_{ij} \nonumber \\ & +
  {G\over7\,c^{5}}\bigg[4S^{(2)}_{a\langle i}M^{(3)}_{j\rangle
      a}+8M^{(2)}_{a\langle i}S^{(3)}_{j\rangle a}
    +17S^{(1)}_{a\langle i}M^{(4)}_{j\rangle a}-3M^{(1)}_{a\langle
      i}S^{(4)}_{j\rangle a}+9S_{a\langle i}M^{(5)}_{j\rangle
      a}\nonumber\\ & \quad\qquad -3M_{a\langle i}S^{(5)}_{j\rangle
      a}-{1\over4}S_{a}M^{(5)}_{ija}-7\varepsilon_{ab\langle
      i}S_{\underline{a}}S^{(4)}_{j\rangle b} +{1\over2}\varepsilon_{ac\langle
      i}\bigg(3M^{(3)}_{\underline{ab}}M^{(3)}_{j\rangle bc}
    +{353\over24}M^{(2)}_{j\rangle bc}M^{(4)}_{ab}\nonumber\\ &
    \quad\qquad -{5\over12}M^{(2)}_{\underline{ab}}M^{(4)}_{j\rangle
      bc}+{113\over8}M^{(1)}_{j\rangle bc}M^{(5)}_{ab}
    -{3\over8}M^{(1)}_{\underline{ab}}M^{(5)}_{j\rangle bc}+{15\over4}M_{j\rangle
      bc}M^{(6)}_{ab} +{3\over8}M_{\underline{ab}}M^{(6)}_{j\rangle
      bc}\bigg)\bigg]\,,\\
V_{ijk}^\text{inst} &= S^{(3)}_{ijk} \nonumber \\ &+ {G\over c^3}
\bigg[ {1\over10}\varepsilon_{ab\langle i}M^{(5)}_{j\underline{a}}M_{k\rangle b}-
  {1\over2}\varepsilon_{ab\langle i}M^{(4)}_{j\underline{a}}M^{(1)}_{k\rangle b} -
  2 S_{\langle i}M^{(4)}_{jk\rangle } \bigg] \nonumber \\ &+
\frac{G}{c^5} \bigg[ \frac{1}{12} M_{a \langle i} S_{jk \rangle
    a}^{(6)} + \frac{1}{12} M_{a \langle i}^{(1)} S_{jk \rangle
    a}^{(5)} + \frac{5}{12} M_{a \langle i}^{(2)} S_{jk \rangle
    a}^{(4)} + \frac{35}{12} M_{a \langle i}^{(4)} S_{jk \rangle
    a}^{(2)} + \frac{49}{12} M_{a \langle i}^{(5)} S_{jk \rangle
    a}^{(1)} \nonumber \\ & \quad\qquad + \frac{19}{12} M_{a \langle
    i}^{(6)} S_{jk \rangle a} + \frac{2}{27} S_{a \langle i} M_{jk
    \rangle a}^{(6)} + \frac{10}{27} S_{a \langle i}^{(1)} M_{jk
    \rangle a}^{(5)} + \frac{2}{27} S_{a \langle i}^{(2)} M_{jk
    \rangle a}^{(4)} + \frac{8}{9} S_{a \langle i}^{(3)} M_{jk \rangle
    a}^{(3)} \nonumber \\ & \quad\qquad - \frac{10}{27} S_{a \langle
    i}^{(4)} M_{jk \rangle a}^{(2)} - \frac{38}{27} S_{a \langle
    i}^{(5)} M_{jk \rangle a}^{(1)} - \frac{2}{3} S_{a \langle
    i}^{(6)} M_{jk \rangle a} - \frac{1}{60} S_a M_{ijka}^{(6)}
  \nonumber \\ & \quad\qquad + \varepsilon_{ab \langle i} \bigg(-
  \frac{1}{180} M_{jk \rangle ac}^{(7)} M_{bc} + \frac{11}{900} M_{jk
    \rangle ac}^{(6)} M_{bc}^{(1)} + \frac{7}{300} M_{jk \rangle
    ac}^{(5)} M_{bc}^{(2)} \nonumber \\ & \quad\qquad\qquad -
  \frac{37}{270} M_{jk \rangle ac}^{(4)} M_{bc}^{(3)} -
  \frac{191}{180} M_{jk \rangle ac}^{(3)} M_{bc}^{(4)} - \frac{65}{36}
  M_{jk \rangle ac}^{(2)} M_{bc}^{(5)} - \frac{367}{300} M_{jk \rangle
    ac}^{(1)} M_{bc}^{(6)} \nonumber \\ & \quad\qquad\qquad -
  \frac{133}{450} M_{jk \rangle ac} M_{bc}^{(7)} + \frac{1}{27} M_{j
    \underline{ac}} M_{k \rangle bc}^{(7)} + \frac{5}{162}
  M_{j\underline{ac}}^{(1)} M_{k \rangle bc}^{(6)} - \frac{5}{162}
  M_{j\underline{ac}}^{(2)} M_{k \rangle bc}^{(5)} \nonumber \\ &
  \nonumber \quad\qquad\qquad - \frac{1}{81} M_{j\underline{ac}}^{(3)}
  M_{k \rangle bc}^{(4)} - \frac{11}{20} S_{jk \rangle b}^{(5)} S_a -
  \frac{88}{45} S_{j \underline{a}} S_{k \rangle b}^{(5)} -
  \frac{40}{9} S_{j\underline{a}}^{(1)} S_{k \rangle b}^{(4)}
  \nonumber \\ &  \quad\qquad\qquad - \frac{32}{9}
  S_{j\underline{a}}^{(2)} S_{k \rangle b}^{(3)} \bigg)\bigg] \,,\\
V_{ijkl}^\text{inst} &= S^{(4)}_{ijkl} \nonumber \\ &+ {G\over
  c^3}\bigg[- {35\over3}S^{(2)}_{\langle
    ij}M^{(3)}_{kl\rangle}-{25\over3}M^{(2)}_{\langle
    ij}S^{(3)}_{kl\rangle} -{65\over6}S^{(1)}_{\langle
    ij}M^{(4)}_{kl\rangle}-{25\over6}M^{(1)}_{\langle
    ij}S^{(4)}_{kl\rangle} -{19\over6}S_{\langle
    ij}M^{(5)}_{kl\rangle}\nonumber\\ & \quad\qquad
  -{11\over6}M_{\langle ij}S^{(5)}_{kl\rangle}-{11\over12}S_{\langle
    i}M^{(5)}_{jkl\rangle} +{1\over6}\varepsilon_{ab\langle
    i}\bigg(-5M^{(3)}_{j\underline{a}}M^{(3)}_{kl\rangle b}
  -{11\over2}M^{(4)}_{j\underline{a}}M^{(2)}_{kl\rangle b}
  -{5\over2}M^{(2)}_{j\underline{a}}M^{(4)}_{kl\rangle b}\nonumber\\ & \quad\qquad
  -{1\over2}M^{(5)}_{j\underline{a}}M^{(1)}_{kl\rangle b}
  +{37\over10}M^{(1)}_{j\underline{a}}M^{(5)}_{kl\rangle b}
  +{3\over10}M^{(6)}_{j\underline{a}}M_{kl\rangle
    b}+{1\over2}M_{j\underline{a}}M^{(6)}_{kl\rangle b}\bigg)\bigg] \,,\\
V_{ijklm}^\text{inst} &= S^{(5)}_{ijklm} \nonumber \\ &+ \frac{G}{c^3}
\bigg[- \frac{3}{2} M_{\langle ij} S_{klm \rangle}^{(6)} -
  \frac{33}{10} M_{\langle ij}^{(1)} S_{klm \rangle}^{(5)} - 12
  M_{\langle ij}^{(2)} S_{klm \rangle}^{(4)} - 27 M_{\langle ij}^{(3)}
  S_{klm \rangle}^{(3)} \nonumber \\ & \quad\qquad - \frac{69}{2}
  M_{\langle ij}^{(4)} S_{klm \rangle}^{(2)} - \frac{39}{2} M_{\langle
    ij}^{(5)} S_{klm \rangle}^{(1)} - \frac{21}{5} M_{\langle
    ij}^{(6)} S_{klm \rangle} - \frac{4}{3} S_{\langle ij} M_{klm
    \rangle}^{(6)} \nonumber \\ & \quad\qquad - \frac{76}{15}
  S_{\langle ij}^{(1)} M_{klm \rangle}^{(5)} - \frac{16}{3} S_{\langle
    ij}^{(2)} M_{klm \rangle}^{(4)} - 8 S_{\langle ij}^{(3)} M_{klm
    \rangle}^{(3)} - \frac{28}{3} S_{\langle ij}^{(4)} M_{klm
    \rangle}^{(2)} \nonumber \\ & \quad\qquad - \frac{20}{3}
  S_{\langle ij}^{(5)} M_{klm \rangle}^{(1)} - \frac{8}{5} S_{\langle
    ij}^{(6)} M_{klm \rangle} - \frac{3}{5} S_{\langle i} M_{jklm
    \rangle}^{(6)} \nonumber \\ & \quad\qquad + \varepsilon_{ab
    \langle i} \Big( \frac{1}{14} M_{j\underline{a}} M_{klm \rangle
    b}^{(7)} + \frac{1}{2} M_{j\underline{a}}^{(1)} M_{klm \rangle
    b}^{(6)} - \frac{3}{5} M_{j\underline{a}}^{(2)} M_{klm \rangle
    b}^{(5)} - \frac{4}{3} M_{j\underline{a}}^{(3)} M_{klm \rangle
    b}^{(4)} \nonumber \\ & \quad\qquad\qquad - \frac{3}{2}
  M_{j\underline{a}}^{(4)} M_{klm \rangle b}^{(3)} - \frac{1}{2}
  M_{j\underline{a}}^{(5)} M_{klm \rangle b}^{(2)} + \frac{1}{35}
  M_{j\underline{a}}^{(7)} M_{klm \rangle b} + \frac{1}{7}
  M_{jk\underline{a}} M_{lm \rangle b}^{(7)} \nonumber \\ &
  \quad\qquad\qquad + \frac{2}{3} M_{jk\underline{a}}^{(1)} M_{lm
    \rangle b}^{(6)} + \frac{4}{3} M_{jk\underline{a}}^{(2)} M_{lm
    \rangle b}^{(5)} + \frac{1}{3} M_{jk \underline{a}}^{(3)} M_{lm
    \rangle b}^{(4)} \Big) \bigg] \,.
\end{align}
\end{subequations}
In the  expressions for the current moments above, the $1/c^5$ terms in
$V_{ijk}^\text{inst}$ and $1/c^3$ terms in $V_{ijklm}^\text{inst}$
are new with this paper. For all higher multipole moments it
  suffices, at this approximation level, to replace $U_L^\text{inst}$ and
  $V_L^\text{inst}$ by the corresponding $M^{(\ell)}_L$ and
  $S^{(\ell)}_L$.

\subsubsection{Tail terms}
\label{sec:tail}

Next we give the contributions due to tails which correspond to
quadratic interactions and arise at (relative) order 1.5PN. They come
from the interaction between the mass monopole $M$ and a non-static
multipole $M_L$ or $S_L$ (with $\ell\geqslant 2$). For these we
dispose of a general formula valid for any $\ell$~\cite{BD92,
  B95}. The following contributions have to be added to any mass and
current multipole moments:
\begin{subequations}
\begin{align}
U_L^\text{tail}(T_R) &= \frac{2 G M}{c^3} \int_{-\infty}^{T_R}
M_L^{(\ell+2)} (\tau) \bigg[ \ln\bigg(\frac{T_R-\tau}{2b}\bigg) +
  \kappa_\ell \bigg] \ud \tau \,,\\ V_L^\text{tail}(T_R) &=
\frac{2 G M}{c^3} \int_{-\infty}^{T_R} S_L^{(\ell+2)} (\tau) \bigg[
  \ln \bigg(\frac{T_R-\tau}{2b} \bigg) + \pi_\ell \bigg] \ud \tau\,.
\end{align}
\end{subequations}
Here the constant $b$ entering the logarithmic kernel is the constant
time scale that has been introduced into the relation~\eqref{TRtr}
between the retarded time $T_R$ in radiative coordinates and the
retarded time $t_r$ in harmonic coordinates. The numerical constants
$\kappa_\ell$ and $\pi_\ell$ are given for general $\ell$, in
\textit{harmonic coordinates}, by
\begin{equation}\label{pikappa}
    \kappa_\ell = \frac{2\ell^2+5\ell+4}{\ell(\ell+1)(\ell+2)} +
    H_{\ell-2}\,,\qquad \pi_\ell = \frac{\ell-1}{\ell(\ell+1)} +
    H_{\ell-1}\,,
\end{equation}
where $H_k \equiv \sum_{j=1}^k \frac{1}{j}$ denotes the $k$-th
harmonic number.

\subsubsection{Tail-of-tail terms}
\label{sec:tail2}

The tails-of-tails arise at relative order 3PN and are due to the cubic
interaction between two monopoles $M$ and one non-static multipole,
\textit{i.e.} $M\times M\times M_L$ or $M\times M\times S_L$. The tail-of-tail
entering the mass quadrupole moment has been already computed in
Ref.~\cite{B98tail}. Those in the mass octupole and current quadrupole
moments, new with this paper, have been obtained by the same
method.\footnote{Tail-of-tail contributions to the gravitational field have
  recently been extensively used in a comparison with gravitational self-force
  results to high post-Newtonian orders~\cite{BFW14a, BFW14b}. The
  corresponding sources to the relaxed Einstein equations can be found in
  Appendix~B of Ref.~\cite{BFW14b}.} The relevant formulas to integrate the
elementary cubic source terms are presented in
Appendix~\ref{app:cubic_integrals}. We find
\begin{subequations}\label{tailtail}
\begin{align}
U_{ij}^\text{tail-tail}(T_R) &= \frac{G^2 M^2}{c^6}
\int_{-\infty}^{T_R} M_{ij}^{(5)}(\tau) \bigg[2 \ln^2
  \bigg(\frac{T_R-\tau}{2b} \bigg) + \frac{11}{3} \ln
  \bigg(\frac{T_R-\tau}{2b} \bigg) \nonumber \\ & \qquad\qquad\quad -
  \frac{214}{105}\ln \bigg(\frac{T_R-\tau}{2\tau_0}\bigg) +
  \frac{124627}{22050}\bigg] \ud \tau \,,\label{tailtailU2}\\
U_{ijk}^\text{tail-tail}(T_R) &= \frac{G^2 M^2}{c^6} 
\int_{-\infty}^{T_R} M_{ijk}^{(6)}(\tau) \bigg[2 \ln^2
  \bigg(\frac{T_R-\tau}{2b} \bigg) + \frac{97}{15} \ln
  \bigg(\frac{T_R-\tau}{2b} \bigg) \nonumber \\ & \qquad\qquad\quad -
  \frac{26}{21} \ln \bigg(\frac{T_R-\tau}{2\tau_0}\bigg) +
  \frac{13283}{4410}\bigg] \ud \tau \,,\label{tailtailU3}\\
V_{ij}^\text{tail-tail}(T_R) &= \frac{G^2 M^2}{c^6} 
\int_{-\infty}^{T_R} J_{ij}^{(5)}(\tau) \bigg[2 \ln^2
  \bigg(\frac{T_R-\tau}{2b} \bigg) + \frac{14}{3} \ln
  \bigg(\frac{T_R-\tau}{2b} \bigg) \nonumber \\ & \qquad\qquad\quad -
  \frac{214}{105} \ln \bigg(\frac{T_R-\tau}{2\tau_0}\bigg) -
  \frac{26254}{11025}\bigg] \ud \tau \,.\label{tailtailV2}
\end{align}
\end{subequations}

Note the appearance of two constant time scales there: (i) The time
scale $b$, which is a pure gauge constant entering the definition of
the particular radiative coordinates used in
Eq.~\eqref{TRtr}. Changing $b$ simply means shifting the origin of
time of the radiative coordinate system $(T,X^i=R N^i)$ with respect
to the harmonic coordinate grid $(t,x^i = r n^i)$, which has clearly
no physical implication. (ii) The time scale $\tau_0=r_0/c$, where
$r_0$ is the regularization constant introduced in the
regulator~\eqref{regulator} of the MPM algorithm;\footnote{When
  studying the case of the mass quadrupole tail-of-tail in
  Ref.~\cite{B98tail}, the choice $b=\tau_0$ was adopted.} the
constant $\tau_0$ cannot be removed by a coordinate transformation,
but it must disappear from the radiative moments (and hence from the
physical waveform) once the \textit{source} moments are explicitly
related to the parameters of the matter source, \textit{e.g.} the
masses, trajectories and velocities of the particles in a binary
system. This has been verified explicitly in the case of the 3PN mass
quadrupole moment in Refs.~\cite{BIJ02, BI04mult}. In
Sec.~\eqref{sec:octupole} below we shall check the cancellation of
$\tau_0$ in the case of the 3PN mass octupole moment as well.

Accordingly, there are two types of logarithms in the kernels of
Eqs.~\eqref{tailtail}: those containing $b$ and those containing
$\tau_0$. The coefficient of the leading logarithm square (which
contains $b$) is always 2.  We also know the coefficient of the
logarithm containing $\tau_0$ in the case of mass-type moments for
general $\ell$. Indeed, the coefficient of $\ln\tau_0$ in the
tails-of-tails associated with mass multipole moments, resulting from
the long computation of the multipole interactions $M\times M\times
M_L$, is given by Eq.~(A6) in Ref.~\cite{BD88} as
\begin{equation}\label{alphaell}
\alpha_\ell = 2\frac{15\ell^4+30\ell^3 +
  28\ell^2+13\ell+24}{\ell(\ell+1)(2\ell+3)(2\ell+1)(2\ell-1)}\,.
\end{equation}
Thus, we have $\alpha_2=214/105$ and $\alpha_3=26/21$ in agreement
with the coefficients displayed in Eqs.~\eqref{tailtailU2}
and~\eqref{tailtailU3}. We shall come back on the significance of this
result in Sec.~\eqref{sec:octupole} when we check that the value
$\alpha_3=26/21$ is fully consistent with our 3PN computation of the
source mass octupole moment $I_{ijk}$ for compact binaries [see
  Eqs.~\eqref{lnr0}--\eqref{tailtaillnr0}]. In Sec.~\ref{sec:resum} we
shall investigate the occurence of the dominant powers of logarithms
for more general tail interactions of the type $M\times \cdots \times
M \times (M_L$ or $S_L)$.

\subsubsection{Memory terms}
\label{sec:mem}

The contributions coming from the non-linear memory effect arise from
quadratic interactions between two radiative moments. They have been
computed in Refs.~\cite{B90, Chr91, WW91, Th92, BD92, B98quad} in the
mass quadrupole moment at the lowest order, which is 2.5PN. More
recently, the 3.5PN corrections beyond leading order have been
obtained for both circular~\cite{F09} and eccentric
orbits~\cite{F11}. Note that the non-linear memory effect does not
enter the current-type multipole moments $V_L$, but only the mass-type
multipole moments $U_L$, hence $V_L^{\text{mem}}=0$. Its contribution
to the mass radiative moments needed for the 3.5PN waveform read
\begin{subequations}
\begin{align}
U_{ij}^{\text{mem}}(T_R) &= \frac{G}{c^5}\int_{-\infty}^{T_R}
\biggl[-\frac{2}{7}M^{(3)}_{a\langle i}(\tau)\,M^{(3)}_{j\rangle
    a}(\tau) \biggr]\ud \tau\nonumber \\ &+ \frac{G}{c^7}
\int_{-\infty }^{T_R} \bigg[- \frac{5}{756} M_{ab}^{(4)} (\tau)
  M_{ijab}^{(4)}(\tau)-\frac{32}{63} S_{a \langle i}^{(3)} (\tau) S_{j
    \rangle a}^{(3)} (\tau) \nonumber \\ &\qquad + \varepsilon_{ab
    \langle i} \bigg( \frac{5}{42} S_{j \rangle bc}^{(4)} (\tau)
  M_{ac}^{(3)} (\tau) - \frac{20}{189} M_{j \rangle bc}^{(4)} (\tau)
  S_{ac}^{(3)} (\tau) \bigg) \bigg] \ud \tau \,,\\
U_{ijk}^\text{mem}(T_R) &= {G\over c^5} \int_{-\infty}^{T_R}
\bigg[-{1\over3}M^{(3)}_{a\langle i} (\tau)M^{(4)}_{jk\rangle a}
(\tau)-{4\over5}\varepsilon_{ab\langle i} M^{(3)}_{j\underline{a}} (\tau)S^{(3)}_{k\rangle
b} (\tau)\bigg]\ud \tau \,,\\ 
U_{ijkl}^\text{mem}(T_R) &= \frac{G}{c^3}\int_{-\infty}^{T_R}
\bigg[{2\over5} M^{(3)}_{\langle ij}(\tau)M^{(3)}_{kl\rangle
  }(\tau)\bigg]\ud\tau \nonumber \\ &+ \frac{G}{c^5}
  \int_{-\infty}^{T_R} \bigg[ \frac{12}{55} M_{a \langle i}^{(4)}
    (\tau) M_{jkl \rangle a}^{(4)}(\tau) - \frac{14}{99} M_{a \langle
      ij}^{(4)}(\tau) M_{kl \rangle a}^{(4)} (\tau) + \frac{32}{45}
    S_{\langle ij}^{(3)}(\tau) S_{kl \rangle}^{(3)} (\tau) \nonumber
    \\ &\qquad + \varepsilon_{ab \langle i} \bigg(- \frac{4}{5} M_{j
      \underline{a}}^{(3)}(\tau) S_{kl \rangle b}^{(4)} (\tau) +
    \frac{32}{45} S_{j \underline{a}}^{(3)}(\tau) M_{kl \rangle
      b}^{(4)}(\tau) \bigg) \bigg] \ud \tau \,,\\ 
U_{ijklm}^\text{mem}(T_R) &= \frac{G}{c^3}\int_{-\infty}^{T_R}
\bigg[{20\over21} M^{(3)}_{\langle
    ij}(\tau)M^{(4)}_{klm\rangle}(\tau)\bigg]\ud\tau \,,\\
U_{ijklmn}^\text{mem}(T_R) &= \frac{G}{c^3} \int_{-\infty}^{T_R}\bigg[
  \frac{5}{7} M_{\langle ijk}^{(4)} (\tau) M_{lmn \rangle}^{(4)}
  (\tau) - \frac{15}{14} M_{\langle ij}^{(4)} (\tau) M_{klmn
    \rangle}^{(4)} (\tau) \bigg] \ud \tau \,.
\end{align}
\end{subequations}
In fact, the non-linear memory terms are known for any multipolar
order $\ell$~\cite{F09}; the above expressions are a particular case
of the general formula given for completeness in
Appendix~\ref{app:memory_modes}, in which we also present the
corresponding modal decomposition of the non-linear memory.

\subsection{Canonical moments in terms of source moments}
\label{sec:cansource}

Adding up all the previous contributions in Eqs.\eqref{UVL} we obtain
the radiative mass and current moments $\{U_L, V_L\}$ as full
functionals of the canonical moments $\{M_L, S_L\}$ consistently with
our 3.5PN goal. However there still remains to relate with that same
precision the canonical moments to the actual sets of source moments
$\{I_L, J_L\}$ and gauge moments $\{W_L, X_L, Y_L, Z_L\}$ as shown
schematically in Eqs.~\eqref{cangen}. Here we present the most
complete up-to-date results, with some repetition (for the sake of
exhaustiveness) with respect to Refs.~\cite{BFIS08, FMBI12}:
\begin{subequations}\label{cansourceMS}
\begin{align}  
M_{ij} &= I_{ij} +\frac{4G}{c^5}
\left[W^{(2)}I_{ij}-W^{(1)}I_{ij}^{(1)}\right] \nonumber\\ &+
\frac{4G}{c^7} \biggl[ - 2 X I_{ij}^{(3)} + \frac{4}{7} I_{a \langle
    i}^{(3)} W_{j \rangle a}^{(1)} + \frac{6}{7} I_{a \langle i}^{(4)}
  W_{j \rangle a} - \frac{1}{7} I_{a \langle i} Y_{j \rangle a}^{(3)}
  - I_{a \langle i}^{(3)} Y_{j \rangle a} + \frac{1}{63} W_a^{(3)}
  I_{ija}^{(1)} \nonumber \\ & - \frac{5}{21} W_a^{(4)} I_{ija} +
  \frac{5}{63} Y_a^{(1)} I_{ija}^{(2)} - \frac{22}{63} Y_a^{(2)}
  I_{ija}^{(1)} - \frac{25}{21} Y_a^{(3)} I_{ija} + 2 W^{(2)}
  W_{ij}^{(1)} \nonumber \\ & + 2 W^{(3)} W_{ij} + 2 W^{(2)} Y_{ij} -
  \frac{4}{3} W_{\langle i} W_{j \rangle}^{(3)} - 4 W_{\langle i} Y_{j
    \rangle}^{(2)} \nonumber \\ & + \varepsilon_{ab \langle i} \bigg(-
  I_{j \rangle a}^{(3)} Z_b + \frac{1}{3} I_{j \rangle a} Z_b^{(3)} +
  \frac{4}{9} J_{j \rangle a} W_b^{(3)} + \frac{8}{9} J_{j \rangle
    a}^{(1)} Y_b^{(1)} - \frac{4}{9} J_{j \rangle a} Y_b^{(2)} \bigg)
  \biggr] + \mathcal{O}\left(\frac{1}{c^8}\right)\,,\\
M_{ijk} &= I_{ijk} + {4G\over
  c^5}\left[W^{(2)}I_{ijk}-W^{(1)}I_{ijk}^{(1)}+3\,I_{\langle
    ij}Y_{k\rangle }^{(1)}\right] +
\mathcal{O}\left(\frac{1}{c^7}\right)\label{M3} \,,\\ 
M_{ijkl} &= I_{ijkl} + \frac{4 G}{c^5} \bigg[- W^{(1)} I_{ijkl}^{(1)} +
  W^{(2)} I_{ijkl} + 4 Y_{\langle i}^{(1)} I_{jkl \rangle} \bigg] +
\mathcal{O}\left(\frac{1}{c^6}\right)\,,\\
S_{ij} &= J_{ij} +{2G\over c^5}\left[\varepsilon_{ab\langle
    i}\left(-I_{j\rangle b}^{(3)}W_{a}-2I_{j\rangle b}Y_{a}^{(2)}
  +I_{j\rangle b}^{(1)}Y_{a}^{(1)}\right)+3J_{\langle i}Y_{j\rangle
  }^{(1)}-2J_{ij}^{(1)}W^{(1)}\right] \nonumber\\& +
\mathcal{O}\left(\frac{1}{c^7}\right)\,,\\
S_{ijk} &= J_{ijk} + \frac{4G}{c^5} \bigg[ - W^{(1)} J_{ijk}^{(1)} +
  \frac{8}{3} Y_{\langle i}^{(1)} J_{jk \rangle} + \varepsilon_{ab
    \langle i} \bigg(- \frac{1}{3} I_{jk \rangle a}^{(1)} Y_b^{(1)} +
  I_{jk \rangle a} Y_b^{(2)} + I_{j\underline{a}}^{(3)} W_{k \rangle
    b} \bigg) \bigg] \nonumber\\& +
\mathcal{O}\left(\frac{1}{c^6}\right)\,.
\end{align}
\end{subequations}
The term $1/c^7$ in $M_{ij}$ was already computed in
Ref.~\cite{FMBI12}, but the terms $1/c^5$ in $M_{ijkl}$ and $S_{ijk}$
are new with the present paper.  Finally, combining the previous
formulas~\eqref{cansourceMS} together with all the results of
Sec.~\ref{sec:radcan} we control the full waveform in terms of the
basic source and gauge moments up to order 3.5PN.

\section{Gravitational-wave octupole modes of compact binaries}
\label{sec:octupole}

In this section we shall present the mass octupole source moment
$I_{ijk}$ of non-spinning compact (point-particle) binaries at 3PN
order for general orbits in the center-of-mass frame, as well as the
associated octupole gravitational-wave modes. These results are part
of our current program to obtain the waveform of compact binaries
complete up to order 3.5PN. The mass octupole moment $I_{ijk}$ is not
conceptually more difficult than the mass quadrupole moment $I_{ij}$,
which has been obtained to order 3PN in Refs.~\cite{BI04mult, BDEI04,
  BDEI05dr} and extended to 3.5PN in Ref.~\cite{FMBI12}. Therefore, we
shall simply present the result of the long calculation, after a short
recapitulation of the method, which is based exactly as in
Refs.~\cite{BI04mult, BDEI04, BDEI05dr} on a preliminary Hadamard type
self-field regularization, followed by dimensional regularization and
renormalization. The more difficult computation of the current
quadrupole moment $J_{ij}$ at the 3PN order will be left for future
work.

\subsection{Dimensional regularization of the mass octupole moment}
\label{sec:DR}

In the first stage of the calculation we obtain the mass octupole
moment by means of the so-called pure-Hadamard-Schwartz (pHS)
regularization to deal with the infinite self-field of the point
particles. The pHS regularization is a specific, minimal Hadamard-type
regularization of integrals, used together with a minimal treatment of
contact ambiguities and Schwartz distributional
derivatives~\cite{BDE04}. It is free of ambiguities but depends on the
usual arbitrary UV regularization length scales $s_A$ ($A=1,2$)
associated with the Hadamard \textit{partie finie} regularization of
integrals that diverge at the locations of the two point
particles~\cite{BFreg}. In addition the constant $r_0$ introduced into
the regulator~\eqref{regulator} is also involved and plays the role of
an IR regularization scale when computing the multipole moments. The
result of this initial calculation thus reads
\begin{equation}\label{pHS}
I_{ijk}^\text{pHS} = I_{ijk}^\text{pHS} \bigl[\overline{\bm{y}}_A,
  s_A, r_0\bigr]\,,
\end{equation}
where we emphasize the dependence on both the UV and IR scales, $s_A$
and $r_0$ respectively, and the functional dependence on the two
trajectories, denoted $\overline{\bm{y}}_A$, with implicit dependence
on the associated coordinate velocities $\overline{\bm{v}}_A =
\ud\overline{\bm{y}}_A/\ud t$.  The trajectories $\overline{\bm{y}}_A$
will later be understood as being ``bare'' trajectories to be
``dressed'' by renormalization [see Eq.~\eqref{shifts}]. Notice that
the initial result~\eqref{pHS} does not constitute by itself a
\textit{physical} solution to the problem. In order to make it
physical within Hadamard's regularization, it must be supplemented by
certain ``\textit{ambiguity terms}''~\cite{BIJ02, BI04mult}. In
dimensional regularization~\cite{Bollini, tHooft}, which is free of
ambiguities and will thus be adopted here, the physical solution is
obtained by augmenting the pHS result with some specific
``\textit{pole part}'' $\propto 1/(d-3)$ in the spatial dimension $d$
considered as a complex number~\cite{BDE04, BDEI04, BDEI05dr}.

Therefore, in the second stage of the calculation, we add to the pHS
result~\eqref{pHS} the so-called ``difference'', which is by
definition what we precisely have to add in order to obtain the
physical result produced by dimensional regularization (DR). The
important point is that the latter difference can be computed purely
\textit{locally}, \textit{i.e.} at the location of the two particles,
in the limit where the dimension tends to three, or equivalently
$\varepsilon\to 0$ with $\varepsilon \equiv d-3$, because it is
determined only by the singular behaviour of integrals in the
neighbourhood of the two singular source points. Moreover, this
difference depends on the Hadamard UV regularization scales $s_A$ as
well as on the DR characteristic parameters, namely $\varepsilon$ and
an arbitrary length scale $\ell_0$ entering Newton's constant in $d$
dimensions, $G^{(d)}=G\ell_0^\varepsilon$. To summarize, the mass
octupole source moment $I_{ijk}^\text{DR}$ in $d=3+\varepsilon$
dimensions is given, in the limit $\varepsilon\to 0$, by
\begin{equation}\label{I3DR}
I_{ijk}^\text{DR}[\overline{\bm{y}}_A, r_0, \varepsilon, \ell_0] =
I_{ijk}^\text{pHS}[\overline{\bm{y}}_A, s_A,r_0] +
\mathcal{D}I_{ijk}[\overline{\bm{y}}_A, s_A, \varepsilon, \ell_0]\,,
\end{equation}
where the second term is made of a polar part $\propto 1/\varepsilon$
plus a finite part contribution $\propto \varepsilon^0$, \textit{i.e.}
is of the type $\mathcal{D}I_{ijk} = \frac{1}{\varepsilon} A_{ijk} +
B_{ijk} + \mathcal{O}\left(\varepsilon\right)$, all terms
$\mathcal{O}(\varepsilon)$ being systematically
neglected.\footnote{One can show that there is only a simple pole at
  order 3PN.} At this stage we check that the two parameters $s_A$
cancel out between the two terms in the right-hand side
of~\eqref{I3DR}, so that $I_{ijk}^\text{DR}$ is free of such arbitrary
UV regularization scales.

Finally, in the third stage of the computation, we
\textit{renormalize} the pole part $\propto 1/\varepsilon$ of the DR
result~\eqref{I3DR} by shifting the particle positions that can be
thought as the bare trajectories
$\overline{\bm{y}}_A\equiv\bm{y}^\text{bare}_A$ into some physical
positions, corresponding to renormalized trajectories
$\bm{y}_A\equiv\bm{y}^\text{renorm}_A$ that will entirely absorb the
pole. The precise \textit{shifts} $\bm{\eta}_A$ of the trajectories,
such that
\begin{equation}\label{shifts}
\overline{\bm{y}}_A = \bm{y}_A + \bm{\eta}_A[\bm{y}_A,
  r'_A,\varepsilon,\ell_0]\,,
\end{equation}
will consist of a pole part followed by a finite part, neglecting a
remainder $\mathcal{O}(\varepsilon)$. These shifts arise at 3PN order
and have been uniquely determined at the same approximation level in
Eqs.~(1.13) and~(6.41)--(6.43) of Ref.~\cite{BDE04}, or Eq.~(6.8) of
Ref.~\cite{BDEI05dr}.  They are precisely those that ensure the
complete equivalence between the Hadamard regularized equations of
motion --- end result of Ref.~\cite{BFeom} --- and the DR equations of
motion obtained in Ref.~\cite{BDE04}. Note that the two UV
regularization scales $r'_A$ entering Eqs.~\eqref{shifts} are
\textit{a priori} different from the parameters $s_A$; they have been
chosen instead to match exactly their counterparts entering the
Hadamard regularized 3PN equations of motion~\cite{BFeom}, in which
they play the role of UV regularization scales in the context of
Hadamard's regularization. Finally our physical, renormalized
(``dressed'') result, which is numerically equal to the original, bare
result, modulo $\mathcal{O}(\varepsilon)$ corrections, \textit{i.e.}
\begin{equation}\label{physDR}
I_{ijk}[\bm{y}_A, r'_A, r_0] \equiv
I^\text{DR}_{ijk}[\overline{\bm{y}}_A, r_0, \varepsilon, \ell_0] +
\mathcal{O}(\varepsilon)\,,
\end{equation}
is finite in the limit where $\varepsilon \to 0$ while keeping the
dressed trajectories $\bm{y}_A$ constant. Using the
link~\eqref{shifts} we can rewrite:
\begin{equation}\label{I3phys}
I_{ijk}[\bm{y}_A, r'_A, r_0] = \lim_{\varepsilon\to
  0}\Bigl\{I_{ijk}^\text{DR}[\bm{y}_A, r_0, \varepsilon, \ell_0] +
\delta_{\bm{\eta}[\bm{y}_A, r'_A,\varepsilon,\ell_0]}I_{ijk}\Bigr\}\,,
\end{equation}
where the modification $\delta_{\bm{\eta}}I_{ijk} = 3 \sum_{A=1,2} m_A
y_A^{\langle i}y_A^j\eta_A^{k\rangle}$ due to the latter shifts
follows from the variation of the Newtonian mass octupole moment
$I^\text{N}_{ijk} = \sum_{A=1,2} m_A y_A^{\langle
  i}y_A^jy_A^{k\rangle}$ (valid in any dimension $d$) induced by
$\bm{\eta}_A$. By construction, the poles $\sim 1/\varepsilon$ cancel
out between the two terms in the right-hand side of Eq.~\eqref{I3phys}
so that the result is indeed finite (and does not depend on $\ell_0$)
in the limit $\varepsilon\to 0$. The final scales it depends upon are
the two UV scales $r'_A$ and the IR scale $r_0$.

Now, the scales $r'_A$ have been shown to be gauge constants,
\textit{i.e.}  removable by a suitable gauge transformation, both in
the equations of motion~\cite{BFeom} and in the radiation
field~\cite{BIJ02, BI04mult}. We shall indeed check that these
constants disappear when we compute the time derivatives of the
octupole moment~\eqref{I3phys} by means of the 3PN equations of motion
for insertion into the waveform. Our final invariant results, namely
the gravitational modes $(3,3)$ and $(3,1)$ obtained in
Sec.~\ref{sec:modes}, are thus independent of $r'_A$.

On the other hand, the dependence on the IR constant $r_0$ offers the
possibility of an interesting consistency check with the expression of
the tails-of-tails for the mass octupole derived previously in
Eq.~\eqref{tailtailU3}. Indeed, as we already mentioned, the
tail-of-tail integrals depend on the constant $\tau_0=r_0/c$, where
$r_0$ is defined by Eq.~\eqref{regulator}, in such a way that it will
exactly cancel the constant $r_0$ coming from the expressions of the
source multipole moments written in terms of the source parameters
(\textit{i.e.} the positions and velocities of the particles). That
such a cancellation between tails-of-tails and source moments actually
occurs has been proved very generally for any isolated matter
system~\cite{B98tail, B98mult}. This has also been explicitly checked
in the case of point particle binaries for the mass quadrupole moment
at 3PN order in Refs.~\cite{BIJ02, BI04mult}. In Sec.~\ref{sec:octCM}
we shall extend this check to the case of the mass octupole moment at
3PN order for general orbits in the center-of-mass frame.

\subsection{The 3PN mass octupole in the center-of-mass frame}
\label{sec:octCM}
 
We have computed the 3PN mass octupole moment~\eqref{I3phys} of two
point masses $\bm{y}_A$ for general orbits in an arbitrary frame. We
then reduced that result to the frame of the center of mass defined by
the nullity of the center-of-mass integral associated with the 3PN
equations of motion, given by Eq.~(2.13) in Ref.~\cite{BI03CM}. An
interesting point about this calculation is that it requires the full
3PN relations between the variables in the center-of-mass frame and
the relative variables, in contrast to what happens for the 3PN mass
quadrupole moment where only the 2PN center-of-mass relations are
needed~\cite{BI04mult}.

The center-of-mass relations for point particle binaries take the
form\footnote{Our notation for point particle binaries is as follows:
  $m_A$ stands for the two masses ($A=1,2$); $m = m_1+m_2$ for the
  total mass; $X_A=m_A/m$ for the two mass fractions; $\nu = X_1X_2$
  for the symmetric mass ratio; $\mu = m\nu$ for the reduced mass;
  $\Delta = X_1-X_2$ for the relative mass difference; $\bm{x}=(x^i) =
  \bm{y}_1 - \bm{y}_2$ and $\bm{v} = (v^i) = \ud \bm{x}/\ud t =
  \bm{v}_1-\bm{v}_2$ for the relative separation and velocity;
  $v^2=\bm{v}^2$ and $\dot{r}=\bm{n}\cdot\bm{v}$, where
  $\bm{n}=\bm{x}/r$ and $r=\vert\bm{x}\vert$.}
\begin{subequations}\label{y12i}
\begin{align}
\bm{y}_1 &= \Big[X_2+\nu\,\Delta\,P\Big] \bm{x} +
\nu\,\Delta\,Q\,\bm{v} \,,\\ \bm{y}_2 &= \Big[-X_1+\nu\,\Delta\, P\Big]
\bm{x} +\nu\,\Delta\,Q\,\bm{v} \,,
\end{align}\end{subequations}
where all the PN corrections are proportional to the symmetric mass
ratio $\nu$ and the mass difference $\Delta=X_1-X_2$. The two
dimensionless coefficients $P$ and $Q$ are given with the full 3PN
precision in Eqs.~(3.13)--(3.14) of Ref.~\cite{BI03CM}.

As an interesting feature the gauge-constants $r'_A$ appear at the 3PN
order, in the coefficient $P$ only, in the form of the particular
combination $r''_0$ defined by
\begin{equation}\label{r0pp}
\Delta\,\ln r''_0 = X_1^2 \ln r'_1 - X_2^2 \ln r'_2
\end{equation}
(see Eq.~(3.15) in Ref.~\cite{BI03CM}), due to the use of the full 3PN
center-of-mass relations for this calculation. However, our final 3PN
mass octupole source moment in the center-of-mass frame
[Eqs.~\eqref{Iijk}--\eqref{ABCDcirc} below] will depend on $r'_A$
through a combination that differs from that of Eq.~\eqref{r0pp},
namely
\begin{equation}\label{r0p}
\ln r'_0 = X_1 \ln r'_1 + X_2 \ln r'_2\,.
\end{equation}

Indeed we have found that the $r'_A$ in the 3PN center-of-mass
relations combine nicely with another combination of these constants
in the 3PN mass octupole moment~\eqref{I3phys} for general orbits,
\textit{i.e.} after applying the required shift of the world
lines~\eqref{shifts} but before the center-of-mass reduction, so that
the final center-of-mass expression of the mass octupole moment
contains the classic combination of these constants as given by
Eq.~\eqref{r0p}. This is perfectly consistent with the fact that the
constants $r'_A$ should \textit{in fine} disappear from
gauge-invariant results such as our final polarization $(3,3)$ and
$(3,1)$ modes~\eqref{hlm}.

The mass octupole moment defined by Eq.~\eqref{I3phys} at the 3PN
order, \textit{i.e.} after the processes of dimensional regularization
and renormalization by shifts of the world lines as reviewed in
Sec.~\ref{sec:DR}, and reduced to the center-of-mass frame using
Eqs.~\eqref{y12i} with full 3PN precision, is finally of the form
\begin{equation}\label{Iijk}
I_{ijk} = - \nu\,m\,\Delta\biggl\{A\,x_{\langle i}x_{j}x_{k\rangle} +
B\,\frac{r}{c}\,v_{\langle i}x_{j}x_{k\rangle} +
C\,\frac{r^2}{c^2}\,v_{\langle i}v_{j}x_{k\rangle} +
D\,\frac{r^3}{c^3}\,v_{\langle i}v_{j}v_{k\rangle} \biggr\}\,.
\end{equation}
The coefficients for general orbits in the center-of-mass frame are
found to be
\begin{subequations}\label{ABCD}
\begin{align}
A &= 1 + \frac{1}{c^{2}} \biggl\{v^2 \Bigl(\frac{5}{6} - \frac{19}{6}
\nu\Bigr) + \frac{G m}{r} \biggl[- \frac{5}{6} + \frac{13}{6}
  \nu\biggr] \biggr\}\nonumber\\ & + \frac{1}{c^{4}} \biggl\{v^4
\Bigl(\frac{257}{440} - \frac{7319}{1320} \nu + \frac{5501}{440}
\nu^2\Bigr) + \frac{G m}{r} \biggl[v^2 \Big( \frac{3853}{1320} -
  \frac{14257}{1320} \nu - \frac{17371}{1320} \nu^2 \Big)\nonumber\\ &
  \qquad + \dot{r}^2 \Big(- \frac{247}{1320} + \frac{531}{440} \nu -
  \frac{1347}{440} \nu^2\Big) \biggr] + \frac{G^2 m^2}{r^2} \biggl[-
  \frac{47}{33} - \frac{1591}{132} \nu + \frac{235}{66}
  \nu^2\biggr]\biggr\} \nonumber\\ & + \frac{1}{c^{5}} \biggl\{ -
\frac{56}{9} \frac{G^2 m^2 \nu \dot{r}}{r^2}\biggr\} \nonumber\\ & +
\frac{1}{c^{6}} \biggl\{v^6 \Bigl(\frac{3235}{6864} -
\frac{7667}{1040} \nu + \frac{10319}{286} \nu^2 - \frac{129707}{2288}
\nu^3\Bigr) \nonumber\\ & \qquad + \frac{G m}{r} \biggl[v^4
  \Big(\frac{11633}{2640} - \frac{93167}{2640} \nu+ \frac{5289}{110}
  \nu^2 + \frac{203299}{2640} \nu^3\Big)\nonumber\\ & \qquad + v^2
  \dot{r}^2 \Big(- \frac{841}{11440} + \frac{2039}{3432} \nu -
  \frac{121393}{11440} \nu^2 + \frac{589981}{17160} \nu^3\Big)
  \nonumber\\ & \qquad + \dot{r}^4 \Big(\frac{1}{208} -
  \frac{1379}{2288} \nu + \frac{44841}{11440} \nu^2 -
  \frac{63699}{11440} \nu^3\Big)\biggr]\nonumber\\ & \qquad +
\frac{G^2 m^2}{r^2} \biggl[v^2 \Big(\frac{40497}{20020} -
  \frac{397027}{8580} \nu + \frac{120069}{2860} \nu^2 -
  \frac{29429}{858} \nu^3\Big)\nonumber\\ & \qquad + \dot{r}^2 \Big(-
  \frac{60413}{120120} + \frac{20801}{1560} \nu - \frac{623351}{17160}
  \nu^2 - \frac{70735}{3432} \nu^3\Big) \biggr]\nonumber\\ & \qquad +
\frac{G^3 m^3}{r^3} \biggl[\frac{4553429}{229320} +
  \frac{902873}{15015} \nu - \frac{31673}{1716} \nu^2 +
  \frac{27085}{5148} \nu^3 - \frac{26}{7} \ln
  \Bigl(\frac{r}{r_0}\Bigr) - 22 \nu \ln
  \Bigl(\frac{r}{r'_{0}}\Bigr)\biggr] \biggr\}\nonumber\\ & +
\mathcal{O}\left(\frac{1}{c^7}\right)\,,\\
B &= \frac{\dot{r}}{c} \biggl\{-1 + 2 \nu\biggr\} \nonumber\\ & +
\frac{\dot{r}}{c^3} \biggl\{v^2 \Bigl(- \frac{13}{22} + \frac{107}{22}
\nu - \frac{102}{11} \nu^2\Bigr) + \frac{G m}{r} \biggl[-
  \frac{2461}{660} + \frac{8689}{660} \nu + \frac{1389}{220}
  \nu^2\biggr]\biggr\}\nonumber\\ & + \frac{1}{c^{4}} \biggl\{ -
\frac{12 G m \nu v^2}{5 r}+\frac{232 G^2 m^2 \nu}{15 r^2}\biggr\}
\nonumber\\ & + \frac{\dot{r}}{c^{5}} \biggl\{ v^4 \Big(-
\frac{2461}{5720} + \frac{37321}{5720} \nu - \frac{34627}{1144} \nu^2
+ \frac{127447}{2860} \nu^3\Big)\nonumber\\ & \qquad + \frac{G m}{r}
\bigg[ v^2 \Big(- \frac{80629}{17160} + \frac{47979}{1144} \nu -
  \frac{167122}{2145} \nu^2 - \frac{267081}{5720}
  \nu^3\Big)\nonumber\\ & \qquad + \dot{r}^2 \Bigl(\frac{5}{572} +
  \frac{1851}{1144} \nu + \frac{3059}{1560} \nu^2 -
  \frac{299171}{17160} \nu^3\Bigr) \bigg]\nonumber\\ & \qquad +
\frac{G^2 m^2}{r^2} \bigg[\frac{229}{9240} + \frac{33459}{440} \nu -
  \frac{1283}{330} \nu^2 + \frac{2287}{264} \nu^3 \bigg]
\biggr\}\nonumber\\ & + \mathcal{O}\left(\frac{1}{c^6}\right)\,,\\
C &= 1 - 2 \nu \nonumber\\ & + \frac{1}{c^{2}} \biggl\{v^2
\Bigl(\frac{61}{110} - \frac{519}{110} \nu + \frac{504}{55}
\nu^2\Bigr) + \dot{r}^2 \Bigl(- \frac{1}{11} + \frac{4}{11} \nu -
\frac{3}{11} \nu^2\Bigr) \nonumber\\ & \qquad + \frac{G m}{r}
\biggl[\frac{1949}{330}+ \frac{62}{165} \nu - \frac{483}{55}
  \nu^2\biggr] \bigg\} \nonumber\\ & + \frac{1}{c^{4}} \biggl\{v^4
\Bigl(\frac{465}{1144} - \frac{35777}{5720} \nu + \frac{3057}{104}
\nu^2 - \frac{25071}{572} \nu^3\Bigr) \nonumber\\ & \qquad + v^2
\dot{r}^2 \Bigl(- \frac{197}{1430} + \frac{1637}{1430} \nu -
\frac{849}{286} \nu^2 + \frac{3063}{1430} \nu^3\Bigr)\nonumber\\ &
\qquad + \frac{G m}{r} \biggl[v^2 \Big(\frac{91379}{17160} -
  \frac{169537}{5720} \nu + \frac{83211}{5720} \nu^2+
  \frac{504721}{8580} \nu^3\Big)\nonumber\\ & \qquad + \dot{r}^2\Big(-
  \frac{3037}{3432} - \frac{125}{104} \nu + \frac{62143}{17160} \nu^2
  + \frac{113089}{8580} \nu^3\Big)\biggr] \nonumber\\ & \qquad +
\frac{G^2 m^2}{r^2} \biggl[- \frac{34515967}{1261260} -
  \frac{75373}{8580} \nu+ \frac{428669}{8580} \nu^2 -
  \frac{62665}{2574} \nu^3 +
  \frac{52}{7}\ln\Bigl(\frac{r}{r_0}\Bigr)\biggr]\biggr\}\nonumber\\ &
+ \mathcal{O}\left(\frac{1}{c^5}\right)\,,\\
D &=\frac{\dot{r}}{c} \biggl\{ \frac{13}{55} - \frac{52}{55} \nu +
\frac{39}{55} \nu^2\biggr\}\nonumber\\ & + \frac{\dot{r}}{c^{3}}
\biggl\{v^2 \Big(\frac{333}{1430} - \frac{3181}{1430} \nu +
\frac{1849}{286} \nu^2 - \frac{7247}{1430} \nu^3\Big) \nonumber\\ &
\qquad + \dot{r}^2 \Big(\frac{112}{2145}- \frac{224}{715} \nu +
\frac{224}{429} \nu^2 - \frac{448}{2145} \nu^3\Big)\nonumber\\ &
\qquad + \frac{G m}{r} \bigg[\frac{26641}{8580} - \frac{8341}{2860}
  \nu - \frac{1655}{156} \nu^2 + \frac{24367}{4290} \nu^3\bigg]
\biggr\} \nonumber\\ & + \mathcal{O}\left(\frac{1}{c^4}\right)\,.
\end{align}
\end{subequations}
Let us also give the result of the reduction to quasi-circular orbits.
Introducing the post-Newtonian parameter $\gamma=G m/(r c^2)$, we have
\begin{subequations}\label{ABCDcirc}
\begin{align}
A^\text{circ} &= 1 -\gamma \nu + \gamma^2 \left( - \frac{139}{330} -
\frac{11923}{660}\nu - \frac{29}{110}\nu^2\right) \nonumber \\ & +
\gamma^3 \left( \frac{1229440}{63063} + \frac{610499}{20020} \nu +
\frac{319823}{17160} \nu^2 - \frac{101}{2340} \nu^3 - \frac{26}{7} \ln
\Bigl(\frac{r}{r_0}\Bigr) - 22 \nu \ln \Bigl(\frac{r}{r'_{0}}\Bigr)
\right) \nonumber \\ & + \mathcal{O}\left(\frac{1}{c^8}\right)\,,\\
B^\text{circ} &= \frac{196}{15}\gamma^2 \nu +
\mathcal{O}\left(\frac{1}{c^6}\right)\,,\\
C^\text{circ} &= 1 - 2\nu +
\gamma \left(\frac{1066}{165} - \frac{1433}{330}\nu + \frac{21}{55}
\nu^2\right) \nonumber \\ & + \gamma^2 \left( - \frac{1130201}{48510}
- \frac{989}{33} \nu + \frac{20359}{330} \nu^2 - \frac{37}{198} \nu^3
+ \frac{52}{7} \ln \Bigl(\frac{r}{r_0}\Bigr)\right) +
\mathcal{O}\left(\frac{1}{c^6}\right)\,,\\
D^\text{circ} &= \mathcal{O}\left(\frac{1}{c^4}\right)\,.
\end{align}
\end{subequations}
We observe that the coefficients $A^\text{circ}$ and $C^\text{circ}$
for quasi-circular orbits are purely conservative, while the other
ones, $B^\text{circ}$ and $D^\text{circ}$, are purely dissipative,
\textit{i.e.} due to radiation reaction. The results~\eqref{ABCDcirc}
extend Eq.~(5.15a) in Ref.~\cite{BFIS08} to 3PN order.

As already mentioned, the IR constant $r_0$ coming from the MPM
regulator~\eqref{regulator} in the 3PN octupole moment is exactly
compensated by the same constant $r_0=c\tau_0$ coming from the kernel
of the mass octupole tail-of-tail integral displayed in
Eq.~\eqref{tailtailU3}. Indeed we can check from
Eqs.~\eqref{Iijk}--\eqref{ABCD} that the octupole moment for general
orbits depends on this constant through the combination
\begin{equation}\label{lnr0}
I_{ijk} = \dots - \frac{26}{21}\frac{G^2 m^2}{c^6}I^{(2)}_{ijk} \ln
r_0 + \mathcal{O}\left(\frac{1}{c^7}\right)\,,
\end{equation}
where we have re-expressed the 3PN coefficient by means of the
Newtonian octupole $I^\text{N}_{ijk} = \sum_A m_A y_A^{\langle
  ijk\rangle}$. The ellipsis denote all the other terms in
Eqs.~\eqref{Iijk}--\eqref{ABCD} which are independent of $r_0$. On the
other hand, Eq.~\eqref{tailtailU3} immediately gives
\begin{equation}\label{tailtaillnr0}
U_{ijk}^\text{tail-tail} = \dots + \frac{26}{21}\frac{G^2
  M^2}{c^6}M^{(5)}_{ijk} \ln r_0 \,,
\end{equation}
which is nicely consistent with the source moment~\eqref{lnr0} and
shows that the octupole tail-of-tail indeed compensates the $r_0$
present in the source octupole, at leading order. Recall that the
coefficient $\alpha_3=26/21$ is a particular case of the general
formula~\eqref{alphaell}.

\subsection{The gravitational-wave octupole modes  $(3,3)$ and $(3,1)$}
\label{sec:modes}

With the 3PN mass octupole source moment in hand, and using all the
non-linear multipole interactions computed in Sec.~\ref{sec:rad}, we
obtain the complete 3PN mass octupole radiative moment $U_{ijk}$. Now,
recall that for non-spinning compact binaries, there is a clean
separation of the modes $(\ell,m)$ into those with $\ell+m$ even,
which depend only on the mass-type moments $U_L$, and those with
$\ell+m$ odd, which depend only on the current-type ones
$V_L$.\footnote{This fact is more generally true for ``planar''
  binaries, whose motion takes place in a fixed orbital plane, which
  is the case of spinning binaries with spins aligned or anti-aligned
  with the orbital angular momentum; see Ref.~\cite{FMBI12} for a
  proof.} From the radiative octupole $U_{ijk}$ we can thus compute
the associated modes $(3,3)$ and $(3,1)$ in the usual spin-weighted
spherical-harmonic decomposition of the waveform~\eqref{gijTT} for
quasi-circular orbits. For $\ell+m$ even we have, adopting the
conventions of Refs.~\cite{BFIS08, FMBI12},
\begin{equation}\label{modes}
h_{\ell m} = - \frac{2G}{R c^{\ell +2}\ell!}
\,\sqrt{\frac{(\ell+1)(\ell+2)}{\ell(\ell-1)}} \,\alpha^L_{\ell
  m}\,U_L\,,
\end{equation}
where the STF tensorial factor $\alpha^L_{\ell m}$ connects the usual
basis of spherical harmonics $Y^{\ell m}$ to the set of STF products
of unit direction vectors $\hat{N}_L$ [see
  Eq.~\eqref{defalphaL}]. Like in Refs.~\cite{BFIS08, FMBI12}, we pose
\begin{equation}\label{modedef}
  h_{\ell m} = \frac{2 G \,m \,\nu \,x}{R \,c^2}
  \,\sqrt{\frac{16\pi}{5}}\, H_{\ell m}\,\mathrm{e}^{-\ui m \, \psi} \,,
\end{equation}
where the post-Newtonian parameter $x=(\frac{G m\omega}{c^3})^{2/3}$
is defined from the orbital frequency of circular motion $\omega$, and
where $\psi$ denotes a particular phase variable related to the actual
orbital phase of the binary, namely $\varphi=\int\omega\ud t$, by
\begin{equation}\label{changephaseomega}
\psi = \varphi - \frac{2 G M \omega}{c^3}
\ln\left(\frac{\omega}{\omega_0}\right) \,,
\end{equation}
the constant frequency $\omega_0$ being directly linked to the time
scale $b$ entering the relation~\eqref{TRtr} between harmonic and
radiative coordinates:
\begin{equation}\label{omega0}
\omega_0=\frac{\mathrm{e}^{\frac{11}{12}-\gamma_\text{E}}}{4b}\,,
\end{equation}
with $\gamma_\text{E}$ denoting the Euler constant. The 1.5PN
logarithmic phase modulation in Eq.~\eqref{changephaseomega}
originates physically from tails that propagate in the far
zone~\cite{BIWW96, ABIQ04}. The mass $M$ therein is the ADM mass; it
\textit{must} include the relevant post-Newtonian corrections, up to
1PN order in the present case (see Eq.~(5.23) in Ref.~\cite{BFIS08}).

Our final results for the gravitational-wave modes $(3,3)$ and $(3,1)$
at order 3.5PN in the waveform for quasi-circular orbits read
\begin{subequations} \label{hlm}\begin{align}
H_{33} &=-\frac{3}{4} \ui \sqrt{\frac{15}{14}} \,\Delta
\bigg[x^{1/2}+x^{3/2} \biggl(-4+2 \nu \biggr)+x^2 \left(3 \pi +\ui
  \Bigl[-\frac{21}{5}+6 \ln (3/2)\Bigr]\right) \nonumber \\ &+x^{5/2}
  \left(\frac{123}{110}-\frac{1838 \nu }{165}+\frac{887 \nu
    ^2}{330}\right)+x^3 \bigg(-12 \pi +\frac{9 \pi \nu }{2} \nonumber
  \\ & \qquad +\ui \Bigl[\frac{84}{5}-24 \ln \left(3/2\right)+\nu
    \Bigl(-\frac{48103}{1215}+9 \ln \left(3/2\right)\Bigr)\Bigr]\bigg)
  \nonumber \\ & + x^{7/2}\biggl(\frac{19388147}{280280} +
  \frac{492}{35} \ln \left(3/2\right) -18\ln^2 (3/2) -
  \frac{78}{7}\gamma_\text{E} + \frac{3}{2} \pi^2 + 6 \ui \pi
  \Big[-\frac{41}{35} + 3 \ln (3/2) \Big] \nonumber \\ & \qquad +
  \frac{\nu}{8} \Big[- \frac{7055}{429} + \frac{41}{8} \pi^2 \Big] -
  \frac{318841}{17160} \nu^2 + \frac{8237}{2860} \nu^3 - \frac{39}{7}
  \ln (16x) \biggr)\bigg] \nonumber \\ &+
\mathcal{O}\left(\frac{1}{c^8}\right)\,,\\
H_{31} &=\frac{\ui \,\Delta}{12 \sqrt{14}} \bigg[x^{1/2}+x^{3/2}
  \left(-\frac{8}{3}-\frac{2 \nu }{3}\right)+x^2 \left(\pi +\ui
  \Big[-\frac{7}{5}-2 \ln 2\Big]\right) \nonumber \\ &+x^{5/2}
  \left(\frac{607}{198}-\frac{136 \nu }{99}-\frac{247
    \nu^2}{198}\right)+x^3 \bigg(-\frac{8 \pi }{3}-\frac{7 \pi \nu
  }{6} \nonumber \\ & \qquad +\ui \Big[\frac{56}{15}+\frac{16 \ln
      2}{3}+\nu \Big(-\frac{1}{15}+\frac{7 \ln 2}{3}\Big)\Big]\bigg)
  \nonumber \\ &+ x^{7/2}\biggl( \frac{10753397}{1513512} - 2 \ln 2
  \Big[ \frac{212}{105} + \ln 2\Big] - \frac{26}{21} \gamma_\text{E} +
  \frac{\pi^2}{6} -2 \ui \pi \Big[ \frac{41}{105} + \ln 2 \Big]
  \nonumber \\ & \qquad + \frac{\nu}{8} \bigg(- \frac{1738843}{19305}
  + \frac{41}{8} \pi^2 \bigg) + \frac{327059}{30888} \nu^2 -
  \frac{17525}{15444} \nu^3 - \frac{13}{21} \ln x \biggr)\bigg]
\nonumber \\ & + \mathcal{O}\left(\frac{1}{c^8}\right)\,.
\end{align}\end{subequations}
This extends Eqs.~(9.4d) and (9.4f) in Ref.~\cite{BFIS08} by one-half
PN order. We have verified the complete agreement in the test-mass
limit $\nu\to 0$ with the corresponding modes computed by black-hole
perturbation techniques and reported in Eqs.~(4.9) of
Refs.~\cite{FI10}. Notice that the latter work uses a phase variable
which differs from our definition $\psi$ given by
Eq.~\eqref{changephaseomega}; in particular it happens to be different
for each mode $(\ell, m)$.\footnote{It is related to ours in the
    test-mass limit ($m_1\to 0$) by
$$\psi^\text{FI}_{\ell m} = \psi + 2 x^{3/2} \left(\gamma_\text{E} +
  \frac{3}{2}\ln \left(4 x\right) - \frac{17}{12} \right) + \psi_{\ell
    m}^{(3\text{PN})}\,,$$
with $\psi_{\ell m}^{(3\text{PN})}$ being defined by Eq.~(4.5) of
Ref.~\cite{FI10}.}

The other modes $(3,2)$ and $(3,0)$ are known at order 3PN but cannot
be computed at order 3.5PN for now; this computation will have to wait
for the completion of the current quadrupole moment $J_{ij}$
(currently known at order 2.5PN~\cite{BFIS08}) up to order 3PN. The
derivation of the 3PN current quadrupole presents new difficulties
with respect to the 3PN mass quadrupole or octupole moments, and will
be left for future work.

\section{Tail-induced resummed waveform}
\label{sec:factor}

\subsection{Resummation of IR logarithms}
\label{sec:resum}

By implementing the MPM algorithm, it has been shown by induction that
the $n$-th post-Minkowskian coefficient in harmonic coordinates in an
expansion when $r\to +\infty$ with $t_r=\text{const}$ involves powers
of $1/r$ and powers of the logarithm $\ln r$ up to $n-1$, so that its
general structure at future null infinity
reads~\cite{BD86}\footnote{Recall that $t_r\equiv t-r$.  In this
  section we pose $G=c=1$.}
\begin{equation}\label{hMPMstruct}
h^{\alpha\beta}_{(n)}(\mathbf{x}, t) = \sum_{\ell=0}^{+\infty}
\hat{n}_L \biggl\{\sum_{1 \leqslant k \leqslant N\atop 0\leqslant p
  \leqslant n-1} \frac{\bigl(\ln r/b\bigr)^p}{r^k}
F^{\alpha\beta}_{L(n)kp}(t_r) + R_{L(n)N}^{\alpha\beta}(r,t_r)
\biggr\} \,.
\end{equation}
Here $h_{(n)}$ stands for the $n$-th order post-Minkowskian
piece of the gravitational field either in the canonical or the
general MPM algorithms reviewed in Sec.~\ref{sec:MPM}. The functions
$F_{L(n)kp}$ are complicated functionals of the canonical moments
$\{M_L, S_L\}$, or the source and gauge moments, $\{I_L, J_L\}$ and
$\{W_L, X_L, Y_L, Z_L\}$ respectively, depending on the chosen
algorithm. The angular part is expressed with the STF products
$\hat{n}_L$ of unit vectors $n^i=x^i/r$. The remainder
$R_{L(n)N}(r,t_r)$ is $\mathcal{O}(1/r^{N-\epsilon})$, with
$0<\epsilon\ll 1$ taking into account the fact that the expansion
involves powers of $\ln r$. Finally the logarithms in
Eq.~\eqref{hMPMstruct} are conveniently rescaled by means of an
arbitrary constant $b$, being understood that the functions
$F_{L(n)kp}$ are themselves dependent on this constant
$b$. Restricting our attention to the leading coefficient of $1/r$ at
infinity we write
\begin{subequations}\label{hnasymp}
\begin{align}
h^{\alpha\beta}_{(n)} &= \frac{1}{r}
\,z^{\alpha\beta}_{(n)}(\mathbf{n}, \ln r, t_r) + \mathcal{O}\Bigl(
\frac{1}{r^{2-\epsilon}}\Bigr) \,,\\\text{with}\quad
z^{\alpha\beta}_{(n)} &= \sum_{\ell} \hat{n}_L \sum_{p=0}^{n-1}
\left(\ln\frac{r}{b}\right)^p F_{L(n)p}^{\alpha\beta}(t_r)\,.
\end{align}
\end{subequations}

We shall refer to the logarithms generated in the far zone expansion
of the metric at infinity as the \textit{IR logarithms}. They have
their root in the famous logarithmic deviation of the retarded cones
$t_r=\text{const}$ in harmonic coordinates with respect to the true
null cones $u=\text{const}$, where $u$ is a null coordinate
  satisfying $g^{\mu\nu}\partial_\mu u \,\partial_\nu u=0$. The IR
logarithms can be removed by a coordinate transformation order by
order in the MPM expansion. We have already seen in
Eqs.~\eqref{TRtr}--\eqref{gijTT} that one can construct radiative
coordinates that are free of any such IR logarithms. Radiative
coordinates are such that $T_R=u+\mathcal{O}(1/R)$.

In this subsection we consider a particular class of IR
logarithms generated by tails. Recall that in the MPM algorithm the IR
logarithms are produced by source terms behaving like $1/r^2$ when
$r\to +\infty$ with $t_r=\text{const}$ (where we include in the
coefficient the usual dependence on powers of $\ln r$). Although
source terms behaving like $1/r^k$ with $k\geqslant 3$ do generate
$1/r$ contributions after application of the retarded integral
[\textit{i.e.}  $\mathop{\mathrm{FP}}_{B=0} \, \Box^{-1}_\mathrm{ret}
  \widetilde{r}^B$ in the notation of Eq.~\eqref{un}], those are in
the form of source-free retarded waves which do not contain IR
logarithms; this is proved by Lemma~7.2 in Ref.~\cite{BD86}.

Using the leading order behaviour $1/r$ of $h_{(n)}$,
Eqs.~\eqref{hnasymp}, we see that IR logarithms can only come from
that part of the gravitational source term $\Lambda$ in
Eq.~\eqref{EFEa} which is quadratic in $h$, say $\Lambda = N(h,h) +
\mathcal{O}(h^3)$ where $N(h,h)$ is bilinear in $h$ as well as its
space-time derivatives $\partial h$ and $\partial^2 h$. Thus IR
logarithms come only from solving the equation
\begin{align} \label{BoxhN}
\Box h^{\alpha\beta}_{(n)} = \sum_{m=1}^{n-1}
N^{\alpha\beta}\bigl(h_{(m)},h_{(n-m)}\bigr) +
\mathcal{O}\left(\frac{1}{r^{3 - \epsilon}}\right)\,.
\end{align}
Writing the Einstein equations in harmonic coordinates in terms of the
gothic metric and using the asymptotic formula $\partial_\mu
h^{\alpha\beta}_{(n)} = - k_\mu \,\partial_t h^{\alpha\beta}_{(n)} +
\mathcal{O}(1/r^{2-\epsilon})$, where $t_r=\text{const}$ and $k_\mu=
(-1, n^i)$ denotes a null Minkowskian vector, yields \cite{B87, BD92}
\begin{equation} \label{eqnIR}
\Box h^{\alpha\beta}_{(n)} = \frac{1}{r^2}\left[ 4M \,\partial_t^2
  z^{\alpha\beta}_{(n-1)} + k^\alpha k^\beta \sigma_{(n)} \right] +
\mathcal{O}\left(\frac{1}{r^{3 - \epsilon}}\right)\,.
\end{equation}

The second term in that expression is due to the re-radiation of
gravitational waves by the stress-energy tensor of gravitational waves
themselves. This term is responsible for the non-linear memory
effect~\cite{B90, Chr91, WW91, Th92, BD92, B98quad, F09, F11} (see
Sec.~\ref{sec:radcan}). The energy density $\sigma_{(n)}$ is
proportional to the total gravitational-wave flux emitted to order $n$
and reads explicitly
\begin{equation} \label{sigman}
\sigma_{(n)} = \frac{1}{2} \sum_{m=1}^{n-1}
\Bigl(\eta_{\mu\rho}\eta_{\nu\sigma} -
\frac{1}{2}\eta_{\mu\nu}\eta_{\rho\sigma}\Bigr)\partial_t
z^{\mu\nu}_{(m)}\partial_t z^{\rho\sigma}_{(n-m)} \,.
\end{equation}
As proved in Ref.~\cite{B87} (see Lemma~2.1 there) the IR logarithms
coming from the second term in Eq.~\eqref{eqnIR} can be removed, order
by order in the MPM iteration, by means of the gauge transformation
with gauge vector
\begin{equation} \label{gaugelambdan}
\lambda^\alpha_{(n)} = \Box^{-1}_\mathrm{ret}
\left[\frac{k^\alpha}{2r^2}\!\int_{-\infty}^{t_r}\!\ud
  \tau\,\sigma_{(n)}(\mathbf{n}, \ln r, \tau)\right] \,.
\end{equation}
As the retarded integral is convergent there is no need to include the
$\mathop{\mathrm{FP}}_{B=0}$ operation.

In this Appendix we shall be interested in the IR logarithms generated
by the tail term associated with backscatter onto the static
space-time curvature generated by the total ADM mass $M$ of the system
--- the first term in the right-hand side of Eq.~\eqref{eqnIR}. The
mass is introduced in the formalism as the constant monopole moment
$I\equiv M$ in the ``canonical'' linearized metric~\eqref{hcan1}. To
prove that $M$ enters the source term at any MPM order $n$ in the way
shown in~\eqref{eqnIR}, we invoke our assumption that the matter
system is stationary before some instant $-\mathcal{T}$ in the past.
Under this assumption the relaxed Einstein equation~\eqref{EFEa} for
the quadratic metric in the stationary epoch $t\leqslant -\mathcal{T}$
tells us that $\Delta h_{(2)}^{\alpha\beta} = \mathcal{O}(1/r^4)$
hence $h_{(2)}^{\alpha\beta} = \mathcal{O}(1/r^2)$. By immediate
recurrence, we conclude that $h_{(n)}^{\alpha\beta} =
\mathcal{O}(1/r^n)$ when $t\leqslant -\mathcal{T}$. From this we infer
that $k_\mu k_\nu z_{(n)}^{\mu\nu} = 0$ at any time for $n\geqslant
2$, since it is constant, due to the asymptotic form of the
harmonic-gauge condition, and vanishes at early time $t\leqslant
-\mathcal{T}$, while we have $k_\mu k_\nu z_{(1)}^{\mu\nu} = - 4
M$. Thus, for any $n\geqslant 2$, the first term of Eq.~\eqref{eqnIR}
comes only from the coupling between the static part of $h_{(1)}$ and
$h_{(n-1)}$.

We shall solve recursively and look for IR logarithms in the equation
\begin{align} \label{eqBoxhIR}
\Box \bar{h}^{\alpha\beta}_{(n)} = \frac{4 M}{r^2} \partial_t^2
\bar{z}^{\alpha\beta}_{(n-1)} + \mathcal{O}\left(\frac{1}{r^{3 -
    \epsilon}}\right)\,.
\end{align}
We add an overbar to emphasize that we are considering an
approximation of the harmonic-coordinate MPM algorithm in which we
neglect all the IR logarithms generated by the second term of
Eq.~\eqref{eqnIR}. The iteration of Eq.~\eqref{eqBoxhIR} is achieved
most easily in the frequency space. The time Fourier transform of a
function $F(t)$ will be denoted by $\tilde{F}(\Omega) \equiv
\mathcal{F}(F)(\Omega)$ with the convention
\begin{equation}\label{Fourier}
\tilde{F}(\Omega) = \int_{-\infty}^{+\infty} \!\! \ud t \,
F(t)\,\mathrm{e}^{\ui \Omega t} \,,\qquad F(t) =
\int_{-\infty}^{+\infty} \!  \frac{\ud \Omega}{2\pi}
\,\tilde{F}(\Omega)\,\mathrm{e}^{-\ui \Omega t} \,.
\end{equation}
The integration of the relevant $1/r^2$ source terms will be achieved
in the Fourier domain with the help of the following formula:
\begin{align} \label{elementaryIR}
 \Box^{-1}_\mathrm{ret} \Bigl[ \frac{\hat{n}_L}{r^2}
   \left(\ln\frac{r}{b}\right)^p \!F(t_r)\Bigr] &=
 \frac{\hat{n}_L}{2(p+1)r}
 \int_{-\infty}^{+\infty}\frac{\ud\Omega}{2\pi}
 \frac{\tilde{F}(\Omega)\,\mathrm{e}^{-\ui \Omega
     t_r}}{(-\ui\Omega)}\biggl[ \gamma_\ell^{(p+1)}(0,\Omega b) -
   \left(\ln\frac{r}{b}\right)^{p+1} \biggr]\nonumber\\ &+
 \mathcal{O}\Big( \frac{1}{r^{2-\epsilon}}\Big) \,,
\end{align}
in which the function of two variables $\gamma_\ell(B,x)$ is defined
by (with $\Gamma$ the Eulerian function) 
\begin{equation}\label{gammaell}
\gamma_\ell(B,x) = \frac{\Gamma(\ell+1+B)\Gamma(1-B)}{(-2\ui
  x)^{B}\Gamma(\ell+1-B)}\,.
\end{equation}
In Eq.~\eqref{elementaryIR} this function is differentiated $(p+1)$
times with respect to the first variable $B$, and then evaluated at
$B=0$ and for $x=\Omega b$, defining thus
\begin{equation}\label{gammaellexpl}
\gamma_\ell^{(p+1)}(0,\Omega b) \equiv
\left(\frac{\partial^{p+1}\gamma_\ell}{\partial B^{p+1}}\right)(0,
\Omega b)\,.
\end{equation}
Again notice that the retarded integral~\eqref{elementaryIR} is
convergent so there is no need to invoke a finite part operation like
in Eq.~\eqref{un}. For the reader's convenience we provide the proof
of the elementary formula~\eqref{elementaryIR} in
Appendix~\ref{app:proof_formula}.

We can now obtain thanks to the elementary
formula~\eqref{elementaryIR}\footnote{Here we consider only the
  retarded integral of the source term, corresponding to the part
  $\bar{u}^{\alpha\beta}_{(n)}$ of the MPM algorithm [see
    Eqs.~\eqref{hgenn}--\eqref{un}], since the part
  $\bar{v}^{\alpha\beta}_{(n)}$ does not contain IR logarithms.} the
leading order waveform $\bar{z}_{(n)}$ starting from the preceding one
$\bar{z}_{(n-1)}$, both having general structures similar to
Eqs.~\eqref{hnasymp}. This gives some recursion relations for the
functions $F_{L(n)p}$ parametrizing their general structures. These
are given in the Fourier domain for any $n\geqslant 2$ by
\begin{subequations}\label{recursion0}\begin{align}
\tilde{F}_{L(n)p}(\Omega) &= \frac{2\ui M \Omega
}{p}\tilde{F}_{L(n-1)p-1}(\Omega) &\text{for $1\leqslant
  p\leqslant n-1$}\,,\\ \tilde{F}_{L(n)0}(\Omega) &= - \sum_{q=0}^{n-2}
\,\frac{2\ui M \Omega }{q+1} \,\gamma_\ell^{(q+1)}(0,\Omega
b)\,\tilde{F}_{L(n-1)q}(\Omega) &\text{for $p=0$}\,.
\end{align}
\end{subequations}
Such recursion formulas are easily iterated with the result that
\begin{subequations}\label{recursion}\begin{align} 
\tilde{F}_{L(n)p} &= \frac{(2\ui M \Omega)^p}{p!}\tilde{F}_{L(n-p)0}
&\text{for $1\leqslant p\leqslant
  n-1$}\,,\label{recursiona}\\\tilde{F}_{L(n)0} &= - \sum_{q=0}^{n-2}
\,\frac{(2\ui M \Omega)^{q+1}}{(q+1)!}  \,\gamma_\ell^{(q+1)}(0,\Omega
b)\,\tilde{F}_{L(n-q-1)0} &\text{for $p=0$}\,.\label{recursionb}
\end{align}
\end{subequations}

These results yield immediately the complete resummed waveform in the
Fourier domain as follows. For convenience we denote by $\tilde{z}_L$
the STF piece with multipolarity $\ell$ in the full waveform. From the
first result~\eqref{recursiona} we then determine
\begin{equation}\label{zLfull}
\tilde{z}^{\alpha\beta}_L(\ln r, \Omega) =
\sum_{n=1}^{+\infty}\tilde{z}^{\alpha\beta}_{L(n)}(\ln r,
\Omega)=\mathrm{e}^{2\ui M \Omega \ln\left(\frac{r}{b}\right)}
\tilde{F}^{\alpha\beta}_{L0}(\Omega)\,,
\end{equation}
in which $\tilde{F}_{L0}(\Omega)$ refers to the logarithmic-free part
of the full waveform and is defined by
$\tilde{F}_{L0}=\sum_{n=1}^{+\infty}\tilde{F}_{L(n)0}$. Next, the
second result~\eqref{recursionb} gives the logarithmic-free part of
the waveform in terms of the linearized approximation which is nicely
factorized out as
\begin{equation}\label{FL0full}
\tilde{F}_{L0}(\Omega) =
\frac{\tilde{F}_{L(1)0}(\Omega)}{\gamma_\ell(2\ui M \Omega,\Omega
  b)}\,,
\end{equation}
where the denominator is made of the function $\gamma_\ell(B,x)$
now evaluated at $B=2\ui M \Omega$ and $x=\Omega b$, and where
we have used the fact that $\gamma_\ell(0,x)=1$. Note that
$\tilde{z}_{L(1)}(\Omega)=\tilde{F}_{L(1)0}(\Omega)$ is the waveform
at the linear order $n=1$. Finally, combining these two findings with
the expression of the function~\eqref{gammaell} we obtain the final
expression for our resummed tail-modified waveform: 
\begin{equation}\label{resumzL}
\bar{z}_L(\ln r, t_r) = \int_{-\infty}^{+\infty} \!\frac{\ud
  \Omega}{2\pi} \,\frac{\Gamma(\ell+1-2\ui M
  \Omega)}{\Gamma(\ell+1+2\ui M \Omega)\Gamma(1-2\ui M
  \Omega)}\,\tilde{z}_{L(1)}(\Omega)\,\mathrm{e}^{- \ui \Omega t_r +
  2\ui M \Omega \ln\left(2 |\Omega| r\right)+M |\Omega|\pi} \,.
\end{equation}
We gladly notice that the scale $b$ has cancelled out from the final
result~\eqref{resumzL}, in agreement with the fact that the iteration
of the equation~\eqref{eqBoxhIR} does not make any reference to an
arbitrary scale such as $b$. However, the dependence on $b$ is
restored in radiative coordinates through the
redefinition~\eqref{TRtr} of the retarded time $T_R$ at future null
infinity, and the phase factor in Eq.~\eqref{resumzL} becomes
$\mathrm{exp}[-\ui \Omega T_R + 2 \ui M \Omega \ln(2 |\Omega| b)+M
  |\Omega| \pi]$. We observe that all powers of $\ln r$ have been
absorbed into $T_R$, but there are still powers of $\ln (2|\Omega| b)$
left. This motivates the change of phase variable introduced in
Eq.~\eqref{changephaseomega}, the constant frequency $\omega_0$
defined in Eq.~\eqref{omega0} being chosen for convenience to minimize
the number of terms in the waveform. Recall also that, as indicated by
an overbar, the resummed waveform~\eqref{resumzL} does not constitute
the complete resummed waveform in harmonic coordinates because we have
systematically neglected the second term in
Eq.~\eqref{eqnIR}. However, it motivates the introduction of
\textit{factorized} resummed waveforms~\cite{DIN09} which we shall
shortly review and complete in the next subsection for the case of the
mass octupole waveform.

In fact, the solution~\eqref{resumzL} can be obtained without MPM
iteration and resummation by computing directly the solution of a
``scattering'' problem,
\begin{align}\label{scattering}
\Box \bar{h} - \frac{4 M}{r} \partial_t^2 \bar{h} = \bar{S} \,,
\end{align}
where the scattering barrier is the usual potential $M/r$, and where $\bar{S}$
is some effective source. To recover the solution~\eqref{resumzL}, it suffices
to impose that $\bar{S}$ decreases like $\mathcal{O}(1/r^{3-\varepsilon})$ at
infinity. The most general solution of~\eqref{scattering} can be constructed
formally by convoluting the source $\bar{S}$ with the retarded Green function
$G_\text{ret}(x-x')$ of the differential operator $\Box - (4
M/r) \partial^2_t$. In Ref.~\cite{AF97}, $G_\text{ret}(x-x')$ is computed
explicitly using standard techniques of scattering theory~\cite{Messiah}. In
the limit $r \to +\infty$ at $t_r=\text{const}$, it is essentially
proportional to the retarded solution of the modified Whittaker equation and
to the normalization coefficient of the solution regular at the origin, for
$\Omega r \to 0$.

\subsection{Application to the octupole resummed waveform}
\label{sec:appli}

We conclude the paper with an application of our above computation of
the 3.5PN accurate modes $h_{33}$ and $h_{31}$ to the
effective-one-body (EOB) approach~\cite{BuonD99, DNorleans} to
analytically blend PN approximants and numerical-relativity (NR)
results. Factorized resummed waveforms were introduced in
Ref.~\cite{DIN09} and consist of a physically motivated product of the
Newtonian waveform, a relativistic correction coming from an effective
source built from the EOB Hamiltonian, the resummed tail effects
linked to propagation on a Schwarzschild background (see
Sec.~\ref{sec:resum}), a residual tail dephasing $\ui \delta_{\ell m}$
(complex), and finally the $\ell$-th power of a residual relativistic
amplitude correction (thus purely real) denoted $\rho_{\ell m}$. The
factorized resummed waveforms achieve better agreement with NR results
than the conventional Taylor expanded PN waveforms~\cite{DIN09, DN09,
  Buon09, FI10}. For $\ell + m$ even (corresponding to even-parity
$\epsilon=0$) we have
\begin{equation}\label{factor}
H_{\ell m} = H_{\ell m}^\text{N}\,S_\text{eff}\,T_{\ell m}\,\mathrm{e}^{\ui
  \delta_{\ell m}}\,\left(\rho_{\ell m}\right)^\ell\,.
\end{equation}
The Newtonian approximation to any even-parity mode reads
\begin{align}\label{HlmN}
H_{\ell m}^\text{N} =& \frac{(-)^{(\ell-m+2)/2}}{2^{\ell+1}
  (\frac{\ell+m}{2})!  (\frac{\ell-m}{2})!(2\ell-1)!!}
\left(\frac{5(\ell+1)(\ell+2)(\ell+m)!(\ell-m)!}{\ell
  (\ell-1)(2\ell+1)}\right)^{1/2} \!\!s_\ell(\nu) \,(\ui\, m)^\ell
\,x^{\ell/2-1}\,,
\end{align}
where we denote $s_\ell(\nu)\equiv X_2^{\ell-1}+(-)^\ell
X_1^{\ell-1}$, see \textit{e.g.} Eq.~(9.5) in Ref.~\cite{BFIS08}. The
effective source $S_\text{eff} = \frac{H_\text{eff}}{\mu c^2}$ is
given at order 3PN by
\begin{align}\label{Seff}
S_\text{eff} &= 1 -\frac{x}{2} \biggl\{ 1 +\left(-\frac{3}{4} -
\frac{1}{3}\nu\right) x + \left(-\frac{27}{8} + \frac{11}{4}\nu\right)
x^2 \nonumber \\ & \quad \quad + \left( -\frac{675}{64} +
\left[\frac{4417}{72} - \frac{205}{96}\pi^2 \right]\nu -
\frac{17}{6}\nu^2 + \frac{\nu^3}{81} \right) x^3 +
\mathcal{O}\left(\frac{1}{c^8}\right)\biggr\}\,.
\end{align}
For $m\geqslant 0$, the leading Schwarzschild tail factor reads
\begin{equation}\label{Tellm}
T_{\ell m} = \frac{\Gamma\left(\ell+1-2\ui
  k_m\right)}{\Gamma(\ell+1)}\,\mathrm{e}^{k_m\left[\pi +
    2\ui\ln\left(2m\omega_0 b\right)\right]}\,,
\end{equation}
in which we denote $k_m \equiv G M m \omega/c^3$, with $M$ being the ADM mass
to be inserted here at 1PN order like in Eq.~\eqref{changephaseomega} (see
Eq.~(5.23) in Ref.~\cite{BFIS08}), and where $\omega_0 b$ is a pure real
number to be found from Eq.~\eqref{omega0}.\footnote{Our definition takes into
  account the modification of the phase given by Eq.~\eqref{changephaseomega}.
  The factor used in Ref.~\cite{DIN09} is thus related to ours by
  $T^\text{DIN}_{\ell m}=T_{\ell m}\mathrm{e}^{2\ui
    k_m\ln(\omega/\omega_0)}$.} Note that the EOB tail factor~\eqref{Tellm} is
motivated by the factorized resummed waveform~\eqref{resumzL}. Indeed, the
dominant Fourier component of the mode $h_{\ell m}$ defined by
Eq.~\eqref{modedef} is obtained by setting $\Omega = m \omega$ in the
stationary phase approximation.

As for the residual dephasings $\delta_{33}$ and $\delta_{31}$, we
find that they do not receive any finite mass correction
$\mathcal{O}(\nu)$ at order 3PN, and are therefore given by the sum of
the expressions~(22) and (24) of Ref.~\cite{DIN09} truncated at order
2.5PN, and the 3PN contributions in the test mass limit computed in
Eqs.~(5.8c) and~(5.8e) of Ref.~\cite{FI10}, \textit{i.e.} up to order
3PN:
\begin{subequations}\label{deltalm}
\begin{align}
\delta_{33} &= \frac{13}{10} y^{3/2} - \frac{80897}{2430} \nu y^{5/2}
+ \frac{39}{7}\pi y^3 +
\mathcal{O}\left(\frac{1}{c^7}\right)\,,\\ \delta_{31} &=
\frac{13}{30} y^{3/2} - \frac{17}{10}\nu y^{5/2} + \frac{13}{21}\pi
y^3 + \mathcal{O}\left(\frac{1}{c^7}\right)\,,
\end{align}\end{subequations}
with $y = (G M \omega/c^3)^{2/3}$, and $M$ is the ADM mass.

Finally the most important inputs we provide in this application are
the finite mass corrections to order 3PN of the amplitude factors
$\rho_{33}$ and $\rho_{31}$ that will form the main blocks in the
factorized resummation of waveforms~\cite{DIN09, FI10, Fuj22PN}. These
are straightforwardly computed from Eqs.~\eqref{hlm}; for completeness
we report here the full 3PN expressions, extending at 3PN order
Eqs.~(52) and~(54) in Ref.~\cite{DIN09}:
\begin{subequations} \label{rholm}\begin{align}
\rho_{33} &= 1+ \left(-\frac{7}{6}+\frac{2}{3}\nu\right)x +
\left(-\frac{6719}{3960}-\frac{1861}{990}\nu
+\frac{149}{330}\nu^2\right)x^2 \nonumber\\ &\qquad + \left(
\frac{3203101567}{227026800} - \frac{26}{7}\gamma_\text{E} -
\frac{13}{7}\ln\left(36 x\right) + \biggl[-\frac{129509}{25740} +
  \frac{41}{192} \pi^2 \biggr] \nu
\right.\nonumber\\&\qquad\qquad\qquad \left. - \frac{274621}{154440}
\nu^2 + \frac{12011}{46332} \nu^3\right)x^3 +
\mathcal{O}\left(\frac{1}{c^8}\right) \,, \\ \rho_{31} &= 1+
\left(-\frac{13}{18}-\frac{2}{9}\nu\right)x +
\left(\frac{101}{7128}-\frac{1685}{1782}\nu -
\frac{829}{1782}\nu^2\right)x^2\nonumber\\ &\qquad +
\left(\frac{11706720301}{6129723600} - \frac{26}{63}\gamma_\text{E} -
\frac{13}{63}\ln\left(4 x\right) + \biggl[-\frac{9688441}{2084940} +
  \frac{41}{192} \pi^2 \biggr] \nu
\right.\nonumber\\&\qquad\qquad\qquad \left. + \frac{174535}{75816}
\nu^2 - \frac{727247}{1250964} \nu^3\right)x^3 +
\mathcal{O}\left(\frac{1}{c^8}\right) \,.
\end{align}\end{subequations}
This completes our application to EOB resummed waveforms.

\acknowledgments We thank the Indo-French collaboration (IFCPAR) under
which a major part of this work has been carried out. B.R.I. is
grateful to IHES, France, and G.F. and L.B. to RRI, India for their
support during the final stages of the project. We also thank Cyril
Denoux for checking the typesetting of the longest equations.

\appendix

\section{Integration of elementary cubic source terms}
\label{app:cubic_integrals}

In order to compute the tail-of-tail contributions to the waveform in
the far zone, we need to control the dominant asymptotic behaviour at
future null infinity of the (finite part of the) retarded integrals of
relevant cubic-source piece, \textit{i.e.}
$\Lambda^{\alpha\beta}_{(3)}$ in the notation of Eq.~\eqref{un}. This
problem is essentially solved in Appendix A of Ref.~\cite{B98tail},
but we shall provide here some more details and additional material.

For generic elementary source terms with multipolarity $\ell$ and
radial dependence $1/r^{k}$ ($k\geqslant 1$) that involve a non-local
integral whose kernel is a Legendre function $Q_m(x)$, we
consider:\footnote{The Legendre function of the second kind $Q_m(x)$
  considered here has a branch cut from $-\infty$ to $1$, and is
  defined by (the first equality being known as Neumann's formula)
\begin{equation}\label{legendre}
Q_m(x) = \frac{1}{2} \int_{-1}^1 \ud z\,\frac{P_m(z)}{x-z} =
\frac{1}{2} P_m (x) \, \mathrm{ln} \left(\frac{x+1}{x-1} \right)-
\sum^{m}_{ j=1} \frac{1}{j} P_{m-j}(x) P_{j-1}(x)\,,
\end{equation}
where $P_m(z)$ is the usual Legendre polynomial whose Rodrigues'
representation reads
\begin{equation}\label{rodrigues}
P_m(z) = \frac{1}{2^m m!} \frac{\ud^m}{\ud z^m}
\Bigl[(z^2-1)^m\Bigr]\,.
\end{equation}
}
\begin{equation}
\Psi^L_{m,\ell,k}= \mathop{\mathrm{FP}}_{B=0} \, \Box^{-1}_\mathrm{ret}
\Big[\widetilde{r}^B \hat{n}_L r^{-k}\int^{+\infty}_1\!\!\!\!  \ud x
  \, Q_m(x) F (t - rx) \Big] \,.
\end{equation}
Depending on the values of $m$, $\ell$ and $k$, the far-zone expansion
of $\Psi^L_{m,\ell,k}$ when $r\to\infty$ (with $t_r=\text{const}$)
takes one of the following forms. For $k=1$ and $m=\ell$ we have
\begin{equation}\label{eq:1lPsiL}
\Psi^L_{\ell,\ell,1} = -{\hat{n}_L\over 8r} \int^{+\infty}_0 \! \!
\!\!\ud\tau F^{(-1)} (t_r - \tau) \biggl[ \ln^2 \Big( \frac{\tau}{2r}
  \Big) + 4 H_\ell \ln \Big( \frac{\tau}{2r} \Big) + 4 H^2_\ell
  ~\biggr] + \mathcal{O}\Big( \frac{1}{r^{2-\epsilon}}\Big) \,,
\end{equation}
where $H_\ell=\sum_{j=1}^\ell \frac{1}{j}$ is the $\ell$-th harmonic
number, and $F^{(-1)}$ denotes the anti-derivative of $F$ that
vanishes at $-\infty$. For $2 \leqslant k \leqslant \ell+2$ and
$k\geqslant \ell +3$ respectively, we have
\begin{subequations}
\begin{align} \label{eq:klPsiL}
& \Psi^L_{m,\ell,k} = - \alpha_{m,\ell,k} ~\frac{\hat{n}_L}{r} F^{(k-3)}
  (t_r) + \mathcal{O}\Big( \frac{1}{r^{2-\epsilon}}\Big) \,, \\ &
  \Psi^L_{m,\ell,k} = -\frac{\hat{n}_L}{r} \int^{+\infty}_0 \!\!\!\!
  \ud\tau F^{(k-2)} (t_r - \tau)\Big[ \beta_{m,\ell,k} \ln \Big(
    \frac{\tau}{2r_0} \Big) + \gamma_{m,\ell,k} \Big] +
  \mathcal{O}\Big( \frac{1}{r^{2-\epsilon}}\Big) \,.
\end{align}
\end{subequations}
The formula~\eqref{eq:1lPsiL} and the explicit expression of
$\alpha_{m,\ell,k}$ for arbitrary $m$, $\ell$, $k$, are obtained in
Ref.~\cite{B98tail}, as well as the explicit expressions of
$\beta_{m,\ell,k}$ and $\gamma_{m,\ell,k}$ for specific values of $k$
and $\ell$.

Let us first recall the derivation of the coefficient
$\alpha_{m,\ell,k}$, which is given for general values of $k$ and
$\ell$ such that $2 \leqslant k \leqslant \ell+2$ by
\begin{equation} \label{eqkmalphal}
\alpha_{m,\ell,k} = \int_1^{+\infty} \!\!\!\! \ud x \, Q_m(x)
\int_{x}^{+\infty} \ud z \, Q_\ell(z)\,\frac{(z-x)^{k-3}}{(k-3)!}\,.
\end{equation}
It is convenient to introduce as intermediate notation the
$(k-2)$-th anti-derivative of $Q_\ell(x)$ that vanishes at
$x=+\infty$, and given for $k\geqslant 3$ by
\begin{equation}\label{antiderQ}
Q^{(-k+2)}_\ell(x) = - \int_{x}^{+\infty}\ud
z\,Q_\ell(z)\,\frac{(x-z)^{k-3}}{(k-3)!} \,.
\end{equation}
For $k=2$ we naturally pose $Q^{(0)}_\ell(x)=Q_\ell(x)$. Recalling
that $Q_\ell(z)$ behaves like $z^{-\ell-1}$ when $z\to+\infty$ we see
that the integral in the right side is convergent when $k \leqslant
\ell+2$. Anti-derivatives of Legendre functions can straightforwardly
be expanded on the basis of Legendre functions themselves by means of
the recurrence relation $(\ud/\ud z)[Q_{\ell+1}(x)- Q_{\ell-1}(x)] =
(2\ell +1) Q_\ell (x)$. This leads to
\begin{align}\label{expandQ}
Q^{(-k+2)}_\ell(x) &= (-)^k \sum_{j=0}^{k-2}
C_{\ell,j}^{k-2}\,Q_{\ell+2j-k+2}(x)\,,\\ \text{with}\quad
C_{\ell,j}^{k-2} &= (-)^j{\genfrac{(}{)}{0pt}{}{k-2}{j}}
\frac{(2\ell+2j-2k+3)!!}{(2\ell+2j+1)!!}(2\ell-2k+4j+5)\,,
\end{align}
where ${\genfrac{(}{)}{0pt}{}{k-2}{j}}$ is the usual binomial
coefficient. Hence the coefficient $\alpha_{m,\ell,k}$ reads
\begin{equation} \label{alphaalt}
\alpha_{m,\ell,k} = (-)^k \int_1^{+\infty} \!\!\!\! \ud x \,
Q_m(x)\,Q^{(-k+2)}_\ell(x) =
\sum_{j=0}^{k-2}C_{\ell,j}^{k-2}\,j_{m,\ell+2j-k+2}\,,
\end{equation}
where the remaining integral is explicitly given by (see
\textit{e.g.}~\cite{GR})\footnote{Notice that
  $\psi(m+1)-\psi(p+1)=H_m-H_p$ and
  $\psi'(m+1)=\frac{\pi^2}{6}-H_{m,2}$, where $\psi(x)$ is the usual
  logarithmic derivative of the Euler gamma function $\Gamma(x)$ and
  $H_{m,2}= \sum_{j=1}^m \frac{1}{j^2}$ is the $m$-th generalized
  harmonic number of order 2.}
\begin{equation}
\label{jmp}
j_{m,p} = \int_1^{+\infty} \ud x\, Q_m(x) Q_p(x) =
\left\{\begin{array}{l}\displaystyle
\frac{H_m-H_p}{(m-p)(m+p+1)}\quad\text{for $m\not=
  p$}\,,\\[0.6cm]\displaystyle
\frac{1}{2m+1}\left(\frac{\pi^2}{6}-H_{m,2} \right)\quad\text{for
  $m=p$}\,.\end{array}\right.
\end{equation}

We next tackle the case of the two other coefficients
$\beta_{m,\ell,k}$ and $\gamma_{m,\ell,k}$, defined for generic values
of $k$ and $\ell$ such that $k\geqslant \ell +3$ as double integrals,
\begin{subequations} \label{eq:coeffs}
\begin{align} \label{eq:kmbetal}
& \beta_{m,\ell,k} = \int_1^{+\infty} \!\!\!\! \ud x \, Q_m(x)
  \,J_{\ell,k}(x) \, , \\ \label{eq:kmgammal} & \gamma_{m,\ell,k} =
  \int_1^{+\infty} \!\!\!\! \ud x\, Q_m(x) \Big[ \bigl(\ln 2
    +H_{k-3}\bigr) \,J_{\ell,k}(x) - K_{\ell,k}(x) \Big] \, ,
\end{align}
\end{subequations}
with the following definitions:
\begin{subequations}\label{eq:kJl}
\begin{align} 
& J_{\ell,k}(x) = \frac{1}{2} \int_{-1}^1 \ud z \, P_\ell(z)
  \frac{(z-x)^{k-3}}{(k-3)!} \, , \\ & K_{\ell,k}(x) = \frac{1}{2}
  \int_{-1}^1 \ud z \, P_\ell(z) \frac{(z-x)^{k-3}}{(k-3)!} \ln
  (x-z)\, .
\end{align}
\end{subequations}
Similarly to Eq.~\eqref{antiderQ} we introduce the $(k-2)$th
anti-derivative of the Legendre polynomial $P_\ell(x)$ that vanishes
at $x=-1$ and is defined by
\begin{equation}\label{antiderP}
P^{(-k+2)}_\ell(x) = \int_{-1}^{x}\ud z\, P_\ell(z)\frac{(x-z)^{k-3}}{(k-3)!}
\,,
\end{equation}
together with $P^{(0)}_\ell(x)=P_\ell(x)$. For the Legendre polynomial
we have exactly the same expansion as in Eqs.~\eqref{expandQ}, namely
\begin{equation}\label{expandP}
P^{(-k+2)}_\ell(x) = (-)^k \sum_{j=0}^{k-2}
C_{\ell,j}^{k-2}\,P_{\ell+2j-k+2}(x)\,.
\end{equation}
We start by writing $J_{\ell,k}(x)$ and $K_{\ell,k}(x)$ in a way
appropriate for future integration over $x$ with some kernel
$Q_m(x)$. Since the following integral is already known \cite{GR},
\begin{equation}\label{intGR}
\int_1^{+\infty} \ud x\, Q_m(x) (x-1)^{\nu} = 2^{\nu}
\frac{[\Gamma(\nu+1)]^2\,\Gamma(m-\nu)}{\Gamma(m+\nu+2)}\,,
\end{equation}
our strategy will consist in expressing the integrals~\eqref{eq:kJl}
as a sum of monomials $(x-1)^\nu$, with $\nu\in \mathbb{N}$ or
$\mathbb{C}$. This is achieved by performing $k-3$ integrations by
parts on the expressions~\eqref{eq:kJl}. Concerning $J_{\ell,k}(x)$
this results in
\begin{equation}\label{Jellk}
J_{\ell,k}(x) = \frac{(-)^{k+1}}{2}\sum_{p=\ell+1}^{k-2}
\frac{P^{(-p)}_\ell(1)}{(k-p-2)!}\,(x-1)^{k-p-2} \,,
\end{equation}
where the coefficients of each of the monomials are built from the
values at 1 of the anti-derivatives of the Legendre
polynomial. Adapting some formulas for the anti-derivatives of the
Legendre polynomial in Ref.~\cite{GR}, we get the values at $z=1$ as
\begin{equation}
\label{Plk1}
P^{(-p)}_\ell(1) = \left\{\begin{array}{lr} 0 &\text{for $1\leqslant
  p\leqslant\ell$}\,,\\[0.4cm]\displaystyle \frac{(-)^\ell 2^{p}
  (p-1)!}{(p+\ell)!(p-\ell-1)!} &\text{for $\ell+1\leqslant
  p$}\,.\end{array}\right.
\end{equation}
Note that the case for $1\leqslant p\leqslant\ell$ is a simple
consequence of Rodrigues' formula~\eqref{rodrigues}; accordingly, we
have written the sum in Eq.~\eqref{Jellk} as starting with
$p=\ell+1$. Finally, the integral~\eqref{eq:kmbetal} over $z$ can be
computed with the help of formula~\eqref{intGR} with $\nu=k-p-2$.

The treatment of $K_{\ell,k}(x)$ is more involved. It is based on the
fact that $\ln(x-z)=\frac{\ud}{\ud B}[(x-z)^B]_{B=0}$ where $B$ is a
convenient complex parameter. The calculation of $K_{\ell,k}(x)$ is
thus similar to that of $J_{\ell,k}(x)$ with one major difference:
After the $k-3$ integrations by parts, there remains a $x$-dependent
factor $(z-x)^B$ in the source of the integral, which becomes $B
(z-x)^{B-1}$ after a last integration by parts. The presence of the
prefactor $B$ prevents the appearance of logarithms $\ln (x-z)$ when
applying the derivative $\ud/\ud B$ and taking the limit $B \to 0$. We
obtain at the end:
\begin{align} \label{eqkJlogl}
 K_{\ell,k}(x) &= \frac{\ud}{\ud
   B}\biggl[\frac{(-)^{k+1}}{2}\sum_{p=\ell+1}^{k-2}
   \frac{P^{(-p)}_\ell(1)}{(k-3)!}
   \frac{\Gamma(B+k-2)}{\Gamma(B+k-p-1)}\,(x-1)^{B+k-p-2}\biggr]_{B=0}
 \nonumber \\ & + \frac{(-)^k}{2} \int_{-1}^1 \ud z\,
 \frac{P^{(-k+2)}_\ell(z)}{z-x} \,.
\end{align}
The last term in Eq.~\eqref{eqkJlogl} is expanded by means of
relation~\eqref{expandP}, whose validity extends to the case $k
\geqslant \ell+3$, provided that we pose $P_{-m-1}(z) \equiv
-P_{m}(z)$ and $(-2m-1)!! \equiv (-)^m/(2m-1)!!$ for any non-negative
integer $m$. Using also Neumann's formula~\eqref{legendre}, we obtain
\begin{equation}\label{lasteq}
\frac{1}{2} \int_{-1}^1 \ud z\, \frac{P^{(-k+2)}_\ell(z)}{z-x} =
(-)^{k+1} \sum_{j=0}^{k-2} C_{\ell,j}^{k-2}\,Q_{\ell+2j-k+2}(z)\,,
\end{equation}
where we have defined $Q_{-m-1}(z) \equiv -Q_{m}(z)$ for
$m\in\mathbb{N}$. Like for $J_{\ell,k}(x)$, the next step consists in
integrating over $x$ each of the monomials $(x-1)^{B+k-p-2}$ by means
of formula~\eqref{intGR}, but this time with a complex parameter
$\nu=B+k-p-2$. After integration we apply the derivative with respect
to $B$ and take the limit $B\to 0$ which is straightforwardly
performed and yields in particular many terms involving the
logarithmic derivative $\psi$ of the Euler gamma function $\Gamma$
[which relates, for integer arguments, to harmonic numbers through the
  equality $\psi(k+1) -\psi(1)= H_k$]. Note that the integration over
$x$ of the last term in~\eqref{eqkJlogl} produces the coefficient
$\alpha_{m,\ell,k}$; compare Eq.~\eqref{lasteq} with
Eqs.~\eqref{expandQ} and \eqref{alphaalt}. However, the coefficient
$j_{m,p}$ defined in Eqs.~\eqref{jmp} must now be extended to the case
where $p$ is negative by posing $j_{m, -|p| -1} = -j_{m, |p|}$.

Finally we are in a position to write down the expressions for
$\beta_{m,\ell,k}$ and $\gamma_{m,\ell,k}$ [besides that given
  by~\eqref{alphaalt} for $\alpha_{m,\ell,k}$] that are effectively
used in the present work:
\begin{subequations}\label{betagammaused}
\begin{align}
\beta_{m,\ell,k} &= (-)^{k + \ell +1} 2^{k-3} \sum_{j=0}^{k-\ell-3}
\frac{(j+\ell)!}{j!(j+2\ell +1)!}  \frac{(m-k+\ell+2 + j)!}{(k-\ell -3
  -j)!}  \frac{[(k-\ell-3-j)!]^2}{(m+k-\ell-2-j)!} \,
,\\ \gamma_{m,\ell,k} &= (-)^{k + \ell} 2^{k-3} \sum_{j=0}^{k-\ell-3}
\frac{(j+\ell)!}{j!(j+2\ell +1)!}  \frac{(m-k+\ell+2 + j)!}{(k-\ell -3
  -j)!}  \frac{[(k-\ell-3-j)!]^2}{(m+k-\ell-2-j)!} \times \nonumber
\\ & \qquad \qquad \qquad \qquad \qquad \times \Bigl(H_{k-\ell-3-j} -
H_{m+\ell + 2 -k + j} - H_{m+k-\ell -2 -j}\Bigr) + \alpha_{m,\ell,k}
\,.\label{gammaused}
\end{align}
\end{subequations}
%

\section{Proof of an elementary integration formula}
\label{app:proof_formula}

We derive the integration formula~\eqref{elementaryIR} used in
Sec.~\ref{sec:resum}. We start with the following formula, valid
through analytic continuation for any $B\in\mathbb{C}$,
\begin{equation}\label{elementaryIRB}
 \Box^{-1}_\mathrm{ret} \biggl[\left(\frac{r}{b}\right)^B
   \frac{\hat{n}_L}{r^{2}} F(t_r)\biggr] = \frac{\hat{n}_L}{2r}
 \int_0^{+\infty}\!\!  \ud\tau F(t_r-\tau) \,\frac{g_\ell(B)
   (\tau/2b)^B- (r/b)^B}{B} + \mathcal{O}\Big(
 \frac{1}{r^{2-\epsilon}}\Big) \,,
\end{equation}
where the function $g_\ell(B)$ is
\begin{equation} \label{gell}
g_\ell(B) =
\frac{\Gamma(\ell+1+B)\Gamma(1-B)}{\Gamma(1+B)\Gamma(\ell+1-B)} \,.
\end{equation}
This formula~\eqref{elementaryIRB} is deduced from Eq.~(A.2) of
Ref.~\cite{B98quad} by performing the change of variable $s=\tau + r$,
expanding the STF derivative operator $\hat{\partial}_L$ (see~(A.15)
in~\cite{B98quad}), and keeping the leading term of order $1/r$. Here
we are working in a neighbourhood of the value of interest $B=0$; in
particular, we suppose $\Re(B) < \epsilon$. We point out that the
elementary integral~\eqref{elementaryIRB} is convergent for $B$ close
to zero. As usual we assume stationarity in the remote past, which
implies in this case that $F(t) = 0$ when $t< -\mathcal{T}$. Thus,
despite the apparent presence of a pole $1/B$, the explicit
expression~\eqref{elementaryIRB} is finite and regular as $B\to
0$. After differentiating $p$ times with respect to $B$, the left-hand
side of~\eqref{elementaryIRB} acquires $p$ powers of the logarithm of
$r/b$, while the right-hand side is straightforwardly evaluated at
$B=0$ with the help of the Leibniz rule. We find
\begin{align} \label{elementaryIRB2}
 \Box^{-1}_\mathrm{ret} \Bigl[ \frac{\hat{n}_L}{r^2}
   \left(\ln\frac{r}{b}\right)^p \!F(t_r)\Bigr] &=
 \frac{\hat{n}_L}{2(p+1)r} \int_{0}^{+\infty} \!\! \ud\tau F(t_r-\tau)
 \left(\frac{\partial^{p+1} }{\partial B^{p+1}}\left[g_\ell(B)
   \left(\frac{\tau}{2b}\right)^B - \left(\frac{r}{b}\right)^B
   \right]\right)_{B=0} \nonumber\\ &+ \mathcal{O}\Big(
 \frac{1}{r^{2-\epsilon}}\Big) \,.
\end{align}
We see that each application of the retarded integral operator
$\Box^{-1}_\mathrm{ret}$ on a source term increases the maximal power
of the logarithms by one unit. Since there is no logarithm in the
linearized part $h_{(1)}$, we infer by recurrence that the $n$th MPM
coefficient $h_{(n)}$ contains logarithms with maximal power $n-1$
(the reasoning is in fact valid for any piece $\sim 1/r^k$ in the
waveform~\cite{BD86}). Finally the formula~\cite{GR}
\begin{equation}
\int_0^{+\infty}\ud \tau\,\tau^B\,\mathrm{e}^{\ui \Omega \tau} =
\frac{\Gamma(B+1)}{(-\ui \Omega)^{B+1}}\,,
\end{equation}
achieves the integration of Eq.~\eqref{elementaryIRB2} in the
frequency space. This yields
\begin{align} \label{intfourier}
 &\Box^{-1}_\mathrm{ret} \Bigl[ \frac{\hat{n}_L}{r^2}
   \left(\ln\frac{r}{b}\right)^p \!F(t_r)\Bigr] = \nonumber\\ 
 &\qquad\quad \frac{\hat{n}_L}{2(p+1)r}
 \int_{-\infty}^{+\infty}\frac{\ud\Omega}{2\pi}
 \frac{\tilde{F}(\Omega)\,\mathrm{e}^{-\ui \Omega
     t_r}}{(-\ui\Omega)}\left(\frac{\partial^{p+1} }{\partial
   B^{p+1}}\left[\frac{g_\ell(B)\Gamma(B+1)}{(-2\ui \Omega b)^B} -
   \left(\frac{r}{b}\right)^B \right]\right)_{B=0} \!\!+
 \mathcal{O}\Big( \frac{1}{r^{2-\epsilon}}\Big) \,,
\end{align}
which coincides with Eq.~\eqref{elementaryIR}.

\section{Modal decomposition of the non-linear memory}
\label{app:memory_modes}

The expressions for the non-linear memory terms given in the text can
be recovered, after appropriate integrations by part, from the general
formula known for any $\ell$~\cite{F09, F11}:
\begin{subequations}\label{eq:memgen}\begin{align} 
U_{L}^\text{mem} &= \mathcal{U}_{L}^\text{mem} + \text{(instantaneous
  contributions)} \,, \\ \text{with}\qquad \mathcal{U}_{L}^\text{mem}
&= \frac{2 c^{\ell-2} (2\ell +1)!!}{(\ell+1)(\ell+2)}
\int_{-\infty}^{T_R} \ud t \int \ud \Omega \,\frac{\ud E}{\ud t\ud
  \Omega} \,\hat{N}_{L} \,, \label{eq:memgenint}
\end{align}
\end{subequations}
where the instantaneous contributions come from the latter
integrations by part. For convenience these instantaneous terms have
been transferred to the instantaneous part of the radiative moments
[see Eqs.~\eqref{UVL}]. They have been included into our explicit
expressions at 3.5PN order in Sec.~\ref{sec:inst}. Recall in addition
that we have $V_{L}^\text{mem} = 0$ for any current moment (at any PN
order).

Here, $\frac{\ud E}{\ud t\ud \Omega} = \frac{R^2c^3}{32\pi G}
\,(\partial_t g^\text{TT}_{ij})^2$ is the gravitational-wave energy
flux per solid angle unit; the other notations are the same as in
Sec.~\ref{sec:rad}. The right-hand side may be put in a more explicit
form by substituting to $g^\text{TT}_{ij}$ its asymptotic
expansion~\eqref{gijTT} and integrating over the solid angle. The last
operation is achieved by means of the useful identity:
\begin{align}
\int\!\!\frac{ \ud \Omega}{4\pi} \hat{n}_{L_1} \hat{n}_{L_2} \hat{n}_L =
\frac{\ell_1!  \ell_2!  \ell!}{(\frac{\ell_1+\ell_2-\ell}{2})!
  (\frac{\ell_1+\ell-\ell_2}{2})!(\frac{\ell_2+\ell-\ell_1}{2})!}
\mathrm{STF}_{k_1 ...k_\ell}
\frac{\delta^{\langle i_1}_{k_1} \!\!\dots \delta^{i_s}_{k_s}\delta^{
    i_{s+1}}_{\langle j_1} \!\!\dots \delta^{i_{\ell_1} \rangle}_{j_{\ell_1-s}}
  \delta^{k_{s+1}}_{j_{\ell_1-s+1}} \!\!\dots
  \delta^{k_\ell}_{j_{\ell_2}\rangle}}{(\ell+\ell_1+\ell_2)!!} \,,
\end{align}
with $s=(\ell_1+\ell-\ell_2)/2$ and $|\ell_1-\ell_2|\leqslant \ell
\leqslant \ell_1+\ell_2$ (hence $s \leqslant \ell$ and $s \leqslant
\ell_1$).

After some combinatorics we find
\begin{align}
\mathcal{U}_{L}^\text{mem} & = \frac{2G}{c^3 (\ell+1) (\ell+2)} \sum_{p, k}
\frac{1}{c^{2p}p! k} \frac{(2\ell +1)!!}{(2p+2\ell+1)!!}  \biggl(
\genfrac{}{}{0pt}{}{\ell}{k-p} \biggr) \nonumber \\ & \quad\times
\int_{-\infty}^{T_R} \ud\tau \biggl[ d^\ell_{pk} \, U^{(1)}_{P\langle
    K-P} U^{(1)}_{L-[K-P]\rangle P} + \frac{e^\ell_{pk}}{c^2} \,
  V^{(1)}_{P\langle K-P} V^{(1)}_{L-[K-P]\rangle P} \nonumber \\ &
  \qquad\qquad\qquad + \frac{f^\ell_{pk}}{c^2} \varepsilon_{ab\langle
    i_1} \, U^{(1)}_{\underline{aP} K-P-1} V^{(1)}_{L-[K-P]\rangle b
    P} \biggr] \,.
\end{align}
Note that this expression is implicit because the radiative moments
$U_L$ in the right-hand side contain themselves a memory
contribution. The coefficient $d^\ell_{pk}$ reads
\begin{equation}
d^\ell_{pk} = k - 4 p \frac{2p+2\ell+1}{2p+\ell-k} +
\frac{2p(p-1)(2p+2\ell+1)(2p+2\ell-1)}{(k-1)(2p+\ell-k)
  (2p+\ell-k-1)}\,,
\end{equation}
if $\text{max}(0, [(5-\ell)/2]) \leqslant p$ and $\text{max}(p,2)
\leqslant k \leqslant \text{min} (p+\ell, 2p + \ell -2)$ (with
$[\cdots]$ denoting the integer part), and $d^\ell_{pk} = 0$
otherwise. The other coefficients are given by
\begin{equation}
e^\ell_{pk} = \frac{4 k (2p+\ell-k)}{(k+1)(2p+\ell-k+1)}
d^\ell_{pk}\,,
\end{equation}
and
\begin{equation}
f^\ell_{pk} = \frac{8 (k-p)}{(2p+\ell-k+2)} \biggl[
  \frac{p(2p+2\ell+1)}{(k-1)(2p+\ell-k)} -1 \biggr]\,,
\end{equation}
if $\text{max}(0, [(4-\ell)/2]) \leqslant p$ and $\text{max}(p+1,2)
\leqslant k \leqslant \text{min} (p+\ell, 2p + \ell -1)$, and
$f^\ell_{pk} = 0$ otherwise.

For completeness, we shall now provide the corresponding formula for
the modal decomposition of the non-linear memory term. The complex
waveform $h \equiv h_+ - \ui h_\times$ can be decomposed onto an
orthonormal basis of functions with spin-weight -2 defined over the
unit sphere. Here we shall choose the set of spin-weighted spherical
harmonics $Y^{\ell m}_{-2}(\Theta,\Phi)$ and use the same conventions
as in Refs.~\cite{BFIS08, FMBI12}:
\begin{equation}\label{eq:hdecomp}
h = \sum^{+\infty}_{\ell=2}\sum^{\ell}_{m=-\ell} h_{\ell
m} \,Y^{\ell m}_{-2}(\Theta,\Phi)\,.
\end{equation}
The coefficients $h_{\ell m}$ may be written in terms of mass and
current type radiative components $U_{\ell m}$ and $V_{\ell m}$
(corresponding to a decomposition in even and odd parity modes in the
case on non-spinning compact binaries):
\begin{equation}\label{eq:inv}
h_{\ell m} = -\frac{G}{\sqrt{2}\,R\,c^{\ell+2}}\left[U_{\ell
m}-\frac{\ui}{c}V_{\ell m}\right]\,.
\end{equation}
Those are related to the STF radiative moments by~\cite{Th80}:
\begin{subequations} \label{eq:UV}
\begin{align}
U_{\ell m} &=
\frac{4}{\ell!}\,\sqrt{\frac{(\ell+1)(\ell+2)}{2\ell(\ell-1)}}
\,\alpha^L_{\ell m}\,U_L\,,\label{eq:U}\\ V_{\ell m} &
=-\frac{8}{\ell!}\,\sqrt{\frac{\ell(\ell+2)}{2(\ell+1)(\ell-1)}}
\,\alpha^L_{\ell m}\,V_L\,,
\end{align}
\end{subequations}
where $\alpha_L^{\ell m}$ is the unique (constant) STF tensor such
that
\begin{equation} \label{defalphaL}
\hat{N}_L = \sum_{m=-\ell}^{\ell} \alpha^L_{\ell m}\,Y^{\ell
  m}(\Theta,\Phi)\,.
\end{equation}

To compute the memory modes $\mathcal{U}_{\ell m}^\text{mem}\propto
\alpha^L_{\ell m} \mathcal{U}_L^\text{mem}$, we insert the angular integral of
Eq.~\eqref{eq:memgenint} into formula~\eqref{eq:U} and contract
$\alpha^L_{\ell m}$ with $\hat{N}_L$ using the
property~\eqref{defalphaL}~\cite{F09}. It remains to integrate a
product of three harmonic functions. After writing them as Wigner
matrices, we obtain an explicit expression for the required integral
by means of standard integration formulas:
\begin{multline}
\int \frac{\ud \Omega}{4\pi} \overline{Y}^{\ell m}(\Theta,\Phi)
Y^{\ell' m'}_{-2}(\Theta,\Phi) \overline{Y}^{\ell''
  m''}_{-2}(\Theta,\Phi) \\= (-)^{m+m'}
\Big(\frac{(2\ell+1)(2\ell'+1)(2\ell''+1)}{(4\pi)^3}\Big)^{1/2} \left(
\begin{array}{ccc}
\ell & \ell' & \ell'' \\
0 & 2 & -2 
\end{array}
\right) \left(
\begin{array}{ccc}
\ell & \ell' & \ell'' \\
-m & m' & -m'' 
\end{array}
\right) \, .
\end{multline}
This yields for the memory mode $(\ell,m)$
\begin{align}
\mathcal{U}_{\ell m}^\text{mem} &= \frac{G}{c^3}
\Big[\frac{(2\ell+1)(\ell-2)!}{8\pi (\ell +2)!} \Big]^{1/2} \! \! \!
\!  \sum_{k \geqslant \text{max}(0,4-\ell)}
\sum_{\ell'=\text{max}([\frac{k+1}{2}],2)}^{\text{min}([\frac{k}{2}]+\ell,k+\ell-2)}
\sum_{m'=
  \text{max}(-\ell',m+\ell'-\ell-k)}^{\text{min}(\ell',m+\ell-\ell'+k)}
\frac{(-)^{m'}}{c^k} \times \nonumber \\ & \qquad \times
\bigl[(2\ell'+1)(2k+2\ell-2\ell'+1)\bigr]^{1/2}
\bigg(\begin{array}{ccc} \ell & \ell' & k + \ell - \ell' \\ 0 & 2 &
  -2 \end{array} \bigg) \bigg(\begin{array}{ccc} \ell & \ell' & k +
  \ell - \ell' \\ -m & m' & m-m' \end{array} \bigg) \times \nonumber
\\ & \qquad \times \Big[ \int_{-\infty}^{T_R} \ud\tau \,
  U^{(1)}_{\ell' m'} \overline{U}^{(1)}_{k+\ell-\ell'\, m'-m} +
  \frac{1}{c^2} \int_{-\infty}^{T_R} \ud\tau \, V^{(1)}_{\ell' m'}
  \overline{V}^{(1)}_{k+\ell-\ell'\, m'-m} \nonumber \\ & \qquad \quad
  + \frac{\ui}{c} \int_{-\infty}^{T_R} \ud\tau \bigl(U^{(1)}_{\ell'
    m'} \overline{V}^{(1)}_{k+\ell-\ell'\, m'-m}- V^{(1)}_{\ell' m'}
  \overline{U}^{(1)}_{k+\ell-\ell'\, m'-m}\bigr) \Big]
\end{align}
where the even-$k$ (odd-$k$) coefficients cancel each other for memory
integrals over products of multipole moments with different
(identical) parities, as one can check explicitly.

\bibliography{ListeRef.bib}

\end{document}